\documentclass[prl,superscriptaddress,twocolumn,preprintnumbers,%
  amsmath,amssymb]{revtex4-2}
\usepackage[utf8]{inputenc}
\usepackage{amsmath,bm}
\usepackage{graphicx,csvsimple,booktabs,epsfig}

\usepackage{color}
\allowdisplaybreaks
\usepackage[dvipsnames]{xcolor}
\usepackage[breaklinks,colorlinks=true]{hyperref}
\usepackage{mathtools}
\usepackage{amsthm,amssymb}
\usepackage{lipsum}
\usepackage{calrsfs}
\usepackage{tikz}
\usepackage{caption}
\usepackage{subcaption}
\usepackage{csvsimple}
\newcommand\myshade{85}
\colorlet{mylinkcolor}{BrickRed}
\colorlet{mycitecolor}{NavyBlue}
\colorlet{myurlcolor}{Aquamarine}
\hypersetup{
  linkcolor  = mylinkcolor!\myshade!black,
  citecolor  = mycitecolor!\myshade!black,
  urlcolor   = myurlcolor!\myshade!black,
  colorlinks = true,
}

\usepackage{url}
\usepackage{xcolor}

\begin{document}
\title{A Method to Assess Granger Causality, Isolation and Autonomy in the Time and Frequency Domains: Theory and Application to Cerebrovascular Variability}

\author{Laura Sparacino}
\affiliation{Department of Engineering, University of Palermo, Italy}
\author{Yuri Antonacci}
\affiliation{Department of Engineering, University of Palermo, Italy}
\author{Chiara Bar\`a}
\affiliation{Department of Engineering, University of Palermo, Italy}
\author{Angela Valenti}
\affiliation{Department of Engineering, University of Palermo, Italy}
\author{Alberto Porta}
\affiliation{Department of Biomedical Sciences for Health, University of Milan, Milan, Italy; Department of Cardiothoracic, Vascular Anesthesia and Intensive Care, IRCCS Policlinico San Donato, San Donato Milanese, Milan, Italy}
\author{Luca Faes}
\affiliation{Department of Engineering, University of Palermo, Italy}

\begin{abstract}
\textit{Objective}: Concepts of Granger causality (GC) and Granger autonomy (GA) are central to assess the dynamics of coupled physiologic processes. While causality measures have been already proposed and largely applied in time and frequency domains, measures quantifying self-dependencies are still limited to the time-domain formulation and lack of a clear spectral representation.
\textit{Methods}: We embed into the classical linear parametric framework for computing GC from a driver random process $X$ to a target process $Y$ a measure of Granger Isolation (GI) quantifying the part of the dynamics of $Y$ not originating from $X$, and a new spectral measure of GA assessing frequency-specific patterns of self-dependencies in $Y$. The framework is formulated in a way such that the full-frequency integration of the spectral GC, GI and GA measures returns the corresponding time-domain measures.
The measures are illustrated in theoretical simulations and  applied to time series of mean arterial pressure and cerebral blood flow velocity obtained in subjects prone to develop postural syncope and healthy controls.
\textit{Results}: simulations show that GI is complementary to GC but not trivially related to it, while GA reflects the regularity of the internal dynamics of the analyzed target process.
In the application to cerebrovascular interactions, spectral GA quantified the physiological response to postural stress of slow cerebral blood flow oscillations, while spectral GC and GI detected an altered response to postural stress in subjects prone to syncope, likely related to impaired cerebral autoregulation. 
\textit{Conclusion and Significance}: The new spectral measures of GI and GA are useful complements to GC for the analysis of interacting oscillatory processes, and detect physiological and pathological responses to postural stress which cannot be traced in the time domain.
The thorough assessment of causality, isolation and autonomy opens new perspectives for the analysis of coupled biological processes in both physiological and clinical investigations.
\end{abstract}
\maketitle

\section{Introduction}
\label{sec:introduction}
In the wide field of Network Physiology, the dynamic activity of a physiological system and the interactions between two systems are typically analyzed measuring synchronous time series from the systems and applying various methods for time series analysis, with the aim of investigating the underling regulatory mechanisms across a variety of physiological states and pathological conditions \cite{lehnertz2020human}. Typical examples of such analyses are the study of the cardiovascular control performed on heart period and arterial pressure (AP) time series \cite{cohen2002short,faes2017information}, the evaluation of cardiorespiratory synchronization assessing how heartbeats are coupled with the breathing activity \cite{schafer1999synchronization, faes2015information}, and the investigation of cerebrovascular interactions from the variability of mean AP (MAP) and cerebral blood flow velocity (CBFV) \cite{panerai1998frequency,bari2016nonlinear}.

The approaches followed to perform the analysis of physiological time series are diverse, spanning from prediction methods or spectral analysis applied to an individual series to measures of correlation and spectral coherence involving pairs of series \cite{akselrod1985hemodynamic,porta2002quantifying}. 
In addition to non-parametric data-driven approaches, the advancement of techniques exploiting dynamic regression models in the time domain \cite{faes2012measuring} and the principle of spectral factorization in the frequency domain \cite{baccala2022partial} have fostered the assessment of \textit{dynamic} and \textit{directed} interactions between coupled time series, offering to physiologists and clinicians new powerful tools for the investigation of causal and oscillatory mechanisms.
The principle of Wiener-Granger causality is undoubtedly driving the efforts in this direction \cite{granger1969investigating, porta2015wiener}, as documented by the constant development of measures like the causal coherence \cite{porta2002quantifying}, the directed coherence (DC, \cite{baccala1998studying}) and measures based on the causality framework proposed by Geweke \cite{geweke1982measurement, geweke1984measures}.
The latter is particularly important, as it realizes a time series decomposition that defines a time-domain GC index and expands it in terms of its frequency content, providing measures with clear information-theoretic interpretation \cite{chicharro2011spectral}. 
This allows to straightforwardly relate between each other measures defined in different domains, also providing frequency-specific interpretations for information-based indexes.
Nevertheless, the link between all the time-domain and spectral measures of GC has not been fully elucidated yet \cite{chicharro2011spectral}, thus leaving room for the investigation of new measures.

In parallel to exploring the concept of GC, the analysis of coupled processes has evolved also studying the role of \textit{autonomous dynamics}, i.e. interactions that occur internally in a process independently of its link with other processes.
In this context, a so-called Granger autonomy (GA) measure has 
been put forward in \cite{seth2010measuring}, based on a previously defined notion of degree of self-determination of a system \cite{boden1996philosophy, bertschinger2008autonomy}. The idea behind the concept is that a target system is autonomous if it is not controlled by external influences but rather it self-determines its states.
In line with these definitions, herein we provide an interpretation of GA as a measure quantifying how much the internal dynamics of a process contribute to determine its predictability more than the dynamics of other processes potentially connected to it. 
Additionally, this measure has been developed also in the context of information theory and applied to gain insight about the physiological mechanisms governing the autonomous dynamics of a process \cite{faes2015information, faes2017information, porta2015conditional, valente2018univariate}.
However, differently from GC, the concept of GA currently lacks of a valid spectral representation, as spectral measures able to identify the autonomous oscillations in a process have not been defined yet.

The present study introduces a framework for the combined assessment of causal interactions and autonomous dynamics in coupled oscillatory processes. Working in the context of bivariate linear regression models, we first embed into the approach for the computation of time-domain and spectral GC \cite{geweke1982measurement,geweke1984measures} a complementary but not redundant measure denoted as “Granger isolation” (GI), which detects the part of the dynamics in a process not originating from the other process. Then, we develop a new spectral measure of GA able to assess autonomous dynamics in a target process; the measure is frequency-specific and such that its integration over all frequencies returns the known time-domain GA measure. The framework is first illustrated in simulations of interacting oscillatory processes, showing the ability of the GC, GI and GA measures to disambiguate independent and interdependent oscillations. 
Then, it is applied to physiological time series of MAP and mean CBFV (MCBFV) measured in healthy controls and in subjects prone to develop postural-related syncope \cite{bari2016nonlinear,faes2013investigating}. The latter is a common clinical problem consisting in a transient loss of consciousness and postural tone with spontaneous recovery, whose underlying patho-physiological mechanisms are still not fully understood.
Here, we evidence the need of computing spectral measures focusing on specific frequency bands with physiological meaning to reveal mechanisms which remain otherwise hidden in a whole-band time-domain analysis, and show the importance of quantifying both causal interactions and autonomous dynamics for a complete assessment of these mechanisms.
The time- and frequency-domain measures of bivariate interactions provided by the framework are collected in the \textit{GICA} Matlab toolbox and available for free download from \\
\small{\url{https://github.com/LauraSparacino/GICA_toolbox}}.

\section{Methods}
This section presents the mathematical formulation of the measures of GC, GI and GA, computed in both the time and frequency domains, modelling the interactions between two stochastic processes via a linear parametric approach.
The approach is grounded on the classical autoregressive (AR) model description of a discrete-time, zero-mean stationary bivariate stochastic process $\textbf{S}_n=[X_{n} Y_{n}]^\intercal$ given by \cite{chicharro2011spectral,faes2012measuring}
\begin{subequations}
\begin{align}
X_n&=\sum_{k=1}^p a_{xx,k}X_{n-k}+a_{xy,k}Y_{n-k}+U_{x|xy,n}, \label{X_ARX}\\
Y_n&=\sum_{k=1}^p a_{yx,k}X_{n-k}+a_{yy,k}Y_{n-k}+U_{y|xy,n}, \label{Y_ARX}
\end{align}
\label{2ARmodel}
\end{subequations}
where $p$ is the model order, defining the maximum lag used to quantify interactions, the coefficients $a$ quantify the time-lagged interactions within and between the two processes, and $U_{x|xy,n}$ and $U_{y|xy,n}$ are uncorrelated white noise processes with variance $\sigma_{x|xy}^2$ and $\sigma_{y|xy}^2$.
The model (\ref{2ARmodel}) is composed by two auto- and cross-regressive (ARX) models whereby each process is regressed both on its own past and on the past of the other process. In compact form, it can be formulated as $\textbf{S}_n=\sum_{k=1}^p \textbf{A}_{k}\textbf{S}_{n-k}+\textbf{U}_n$, where
$\textbf{A}_{k}$ is a 2x2 coefficient matrix containing $a_{xx,k}$ and $a_{xy,k}$ in the first row and $a_{yx,k}$ and $a_{yy,k}$ in the second row, and $\textbf{U}_n=[U_{x|xy,n} U_{y|xy,n}]^\intercal$.


\subsection{Granger Causality and Granger Isolation}
Given the stochastic processes $X$ and $Y$, the very popular concept of GC is formalized quantifying the improvement in predictability that the past states of a putative driver process (say $X$) bring to the present state of the target process (say $Y$) above and beyond the predictability brought by the past states of the target itself \cite{granger1969investigating}. To implement this concept in the context of linear regression models, the present state of the target, $Y_n$, is described first from the past of both $X$ and $Y$ through the so-called \textit{full} model (\ref{Y_ARX}), and then from the past of $Y$ only through the \textit{restricted} AR model
\begin{equation}
Y_n=\sum_{k=1}^\infty b_{yy,k}Y_{n-k}+U_{y|y,n}, \label{ARmodel}
\end{equation}
where $b_{yy,k}$ are AR coefficients and $U_{y|y,n}$ is a white noise process with variance $\sigma_{y|y}^2$; note that the order of the restricted AR model is theoretically infinite, and see the supplemental material (\textit{Sect. S1.A,C}) for the identification of its parameters.
The predictability improvement is typically quantified by the logarithmic measure of GC from $X$ to $Y$ \cite{geweke1982measurement}
\begin{equation}
F_{X\to{Y}}=\ln{\frac{\sigma_{y|y}^2}{\sigma_{y|xy}^2}}, \label{GCtime}
\end{equation}
which takes values going from $F_{X\to{Y}}=0$, when the full model does not yield any predictability improvement  ($\sigma_{y|xy}^2=\sigma_{y|y}^2$), to $F_{X\to{Y}}\to\infty$, when the full model explains completely the target dynamics ($\sigma_{y|xy}^2\to 0$). The logarithmic GC measure (\ref{GCtime}) has an information-theoretic meaning as
for Gaussian processes it is equivalent, up to a factor 2, to the  transfer entropy measure \cite{barnett2009granger,faes2017information, faes2015information}.


To analyze causal interactions in the frequency domain, the model coefficients can be first represented through the \textit{Z}-transform of (\ref{2ARmodel}), yielding $\textbf{S}(z)=\textbf{H}(z)\textbf{U}(z)$, where $\textbf{H}(z)=[\textbf{I}-\sum_{k=1}^p \textbf{A}_{k}z^{-k}]^{-1}$ is the 2x2 transfer matrix, being $\textbf{I}$ the 2x2 identity matrix. Computing $\textbf{H}(z)$ on the unit circle in the complex plane ($\textbf{H}(\bar{f})=\textbf{H}(z)|_{z=e^{j2\pi{\bar{f}}}}$, where $\bar{f} = \frac{f}{f_s} \in{[-0.5,0.5]}$ is the normalized frequency, with $f$ the frequency and $f_s$ the sampling frequency), the 2x2 power spectral density (PSD) matrix of the bivariate process is $\textbf{P}(\bar{f})=\textbf{H}(\bar{f})\mathbf{\Sigma}\textbf{H}^*(\bar{f})$, where $\mathbf{\Sigma}=\mathbb{E}[\textbf{U}_n \textbf{U}_n^\intercal]$ is the covariance of $\textbf{U}_n$ and $^*$ stands for Hermitian transpose \cite{faes2012measuring}. This matrix contains the PSDs of $X$ and $Y$ and the cross-PSDs between $X$ and $Y$ as diagonal and off-diagonal elements, respectively.
Under the hypothesis of strict causality leading to diagonality of $\mathbf{\Sigma}$ \cite{faes2012measuring,ding2006granger}, the PSD of the target process $Y$ can be factorized as
\begin{equation}
P_{Y}(\bar{f})=\sigma_{x|xy}^2 |H_{yx}(\bar{f})|^2 + \sigma_{y|xy}^2 |H_{yy}(\bar{f})|^2. \label{spectraldecomposition}
\end{equation}
From this factorization, the squared directed coherence (DC) from the driver $X$ to the target $Y$ is defined as \cite{baccala1998studying}
\begin{equation}
|\gamma_{YX}(\bar{f})|^2 = \frac{\sigma_{x|xy}^2 |H_{yx}(\bar{f})|^2}{\sigma_{x|xy}^2 |H_{yx}(\bar{f})|^2 + \sigma_{y|xy}^2 |H_{yy}(\bar{f})|^2}, \label{DC}
\end{equation}
measuring the coupling strength from $X$ to $Y$ as the normalized portion of $P_{Y}(\bar{f})$ due to the driver process $X$. Similarly to (\ref{DC}), it is possible to define the normalized portion of $P_{Y}(\bar{f})$ which arises from the target process $Y$ itself as 
\begin{equation}
|\gamma_{YY}(\bar{f})|^2 = \frac{\sigma_{y|xy}^2 |H_{yy}(\bar{f})|^2}{\sigma_{x|xy}^2 |H_{yx}(\bar{f})|^2 + \sigma_{y|xy}^2 |H_{yy}(\bar{f})|^2}. \label{DC_diag}
\end{equation}
The DC measures (\ref{DC}) and (\ref{DC_diag}) allow to decompose the PSD of the target process as $P_{Y}(\bar{f})=P_{Y|X}(\bar{f})+P_{Y|Y}(\bar{f})$:  
$P_{Y|X}(\bar{f})=|\gamma_{YX}(\bar{f})|^2 P_{Y}(\bar{f})$ is the part of $P_{Y}(\bar{f})$ due to $X$, which is usually referred to as the \textit{causal} part of the target spectrum; $P_{Y|Y}(\bar{f})=|\gamma_{YY}(\bar{f})|^2 P_{Y}(\bar{f})$ measures the part of $P_{Y}(\bar{f})$ due to the process $Y$ itself, which may be thus referred to as the \textit{isolated} part of the target spectrum.

Importantly, the DC defined in (\ref{DC}) can be regarded as a measure of GC from $X$ to $Y$ thanks to its relation with the logarithmic spectral measure of GC defined by Geweke \cite{geweke1982measurement}
\begin{equation}
f_{X\to{Y}}(\bar{f})=\ln{\frac{P_Y(\bar{f})}{\sigma_{y|xy}^2 |H_{yy}(\bar{f})|^2}}, \label{GewekeGCfreq} 
\end{equation}
which is linked to the time-domain GC measure (\ref{GCtime}) by the spectral integration property
\begin{equation}
F_{X\to{Y}} = 2 \int_0^{\frac{1}{2}} f_{X\to{Y}}(\bar{f})\,d\bar{f}. \label{GCspectint}
\end{equation}
In fact, combining (\ref{spectraldecomposition}), (\ref{DC}) and (\ref{GewekeGCfreq}) one can easily show that the DC and the spectral GC are linked by the relation $f_{X\to{Y}}(\bar{f})= - \ln({1-|\gamma_{YX}(\bar{f})|^2})$ 
\cite{geweke1982measurement, chicharro2011spectral, faes2012measuring}.
In analogy with these derivations, here we propose a new spectral logarithmic measure of \textit{Granger isolation} (GI) of $Y$ linked to the isolated part of the target spectrum and hence defined as
\begin{equation}
f_{Y}(\bar{f})=\ln{\frac{P_{Y}(\bar{f})}{\sigma_{x|xy}^2 |H_{yx}(\bar{f})|^2}} = - \ln({1-|\gamma_{YY}(\bar{f})|^2}). \label{nGCfreq}
\end{equation}
Moreover, following (\ref{GCspectint}) we provide a new time-domain measure of GI integrating the spectral measure in (\ref{nGCfreq}) over all frequencies:
\begin{equation}
F_{Y} = 2 \int_0^{\frac{1}{2}} f_{Y}(\bar{f})\,d\bar{f} \label{GnCtime}.
\end{equation}

Intuitively, one might think that the GI measure (\ref{GnCtime}) reflects the concept of Granger Autonomy \cite{seth2010measuring}, given that it is derived from the isolated (non-causal) part of the target spectrum.
However, we will show that the GI behaves differently than the known GA measure defined from the error variances of linear regression models \cite{seth2010measuring, faes2017information, porta2015conditional, faes2015information}; mathematical details are provided in the following section, together with the definition of a new spectral measure of GA which satisfies the spectral integration property.


\subsection{Granger Autonomy}
In analogy with GC, the concept of GA is formalized for a bivariate process assessing the predictability improvement brought to the present state of the target $Y$ by its own past states above and beyond the predictability brought by the past states of the driver $X$ \cite{seth2010measuring}. Operationally, GA is quantified comparing the full model (\ref{Y_ARX}) with a restricted cross-regressive (X) model whereby $Y_n$ is described only from the past of $X$:
\begin{equation}
Y_n=\sum_{k=1}^\infty b_{yx,k}X_{n-k}+U_{y|x,n}
\label{Y_X}
\end{equation}
where $b_{yx,k}$ are cross-regression coefficients and $U_{y|x}$ is an innovation process with variance $\sigma_{y|x}^2$. The derivation of the parameters of the restricted model (\ref{Y_X}) is reported in \textit{Sect. S1.B,C} of the supplemental material. Then, in analogy to (\ref{GCtime}), the predictability improvement is quantified by the logarithmic measure of GA given by \cite{seth2010measuring}
\begin{equation}
A_{Y}=\ln \frac{\sigma_{y|x}^2}{\sigma_{y|xy}^2}, \label{GAtime}
\end{equation}
which quantifies the strength of the autonomous dynamics of $Y$ comparing the error variances of the models (\ref{Y_ARX}) and (\ref{Y_X}). In the case of Gaussian processes, the GA measure (\ref{GAtime}) is equivalent, up to a factor 2, to the information-theoretic measure of conditional self-entropy \cite{faes2017information, porta2015conditional, faes2015information}.

Now we derive the spectral representation that leads to the definition of our new measure of GA in the frequency domain. To this end, we first describe the bivariate AR model formed by (\ref{X_ARX}) and (\ref{Y_X}) in the \textit{Z} domain as $\textbf{S}(z)=\textbf{G}(z)\textbf{W}(z)$, where $\textbf{W}(z)$ is the Z-transform of the noise vector $\textbf{W}_n=[U_{x|xy,n} U_{y|x,n}]^\intercal$ and the 2$\times$2 transfer matrix is
\begin{equation}
    \textbf{G}(z) = \begin{bmatrix}
                                G_{xx}(z) & G_{xy}(z) \\
                                G_{yx}(z) & G_{yy}(z)\\
                            \end{bmatrix}
    = \begin{bmatrix}
                                1-A_{xx}(z) & -A_{xy}(z) \\
                                -B_{yx}(z) & 1\\
                            \end{bmatrix}^{-1},
\label{trasfermatrixrestricted}
\end{equation}
with $A_{xx}(z)=\sum_{k=1}^p a_{xx,k}z^{-k}$, $A_{xy}(z)=\sum_{k=1}^p a_{xy,k}z^{-k}$, $B_{yx}(z)=\sum_{k=1}^p b_{yx,k}z^{-k}$. Computing $\textbf{G}(z)$ on the unit circle of the complex plane ($z=e^{j2\pi{\bar{f}}}$) yields the 2$\times$2 complex transfer function in the frequency domain, $\textbf{G}(\bar{f})$.

At this point, we note that in (\ref{Y_X}) the removal of the predictable autonomous dynamics of the target process makes them likely to be contained in the residual $U_{y|x}$, and thus not modelled by the element $G_{yy}(\bar{f})$ of the transfer function matrix $\textbf{G}(\bar{f})$. Since in (\ref{Y_ARX}) these autonomous dynamics are instead modelled by $H_{yy}(\bar{f})$, they can be emphasized comparing the two transfer functions of the full and restricted models. Accordingly, we propose to assess the strength and frequency-specific location of the target internal dynamics through the spectral function
\begin{equation}
\bar{a}_{Y}(\bar{f})=\ln \frac{|{H}_{yy}(\bar{f})|^2}{|{G}_{yy}(\bar{f})|^2} \label{GAbarfreq},
\end{equation}
which captures the balance between the transfer of information within the target quantified when the self-dependencies are modelled and when they are not. We expect that stronger internal dynamics at the frequency $\bar{f}$ are reflected by higher values of $|{H}_{yy}(\bar{f})|^2$ compared with $|{G}_{yy}(\bar{f})|^2$, and thus to higher values of $\bar{a}_{Y}(\bar{f})$.
However, since the full-frequency integral of both $\ln |{H}_{yy}(\bar{f})|^2$ and $\ln |{G}_{yy}(\bar{f})|^2$ is null \cite{rozanov1967stationary}, we have that $2 \int_0^{\frac{1}{2}} \bar{a}_{Y}(\bar{f})d\bar{f}=0$, and thus $\bar{a}_{Y}(\bar{f})$ will take negative values at some frequencies, and its full-frequency integral will not return the time-domain GA. To counteract these issues, we introduce the spectral GA measure defined as
\begin{equation}
a_{Y}(\bar{f})=\ln \frac{\sigma_{y|x}^2|{H}_{yy}(\bar{f})|^2}{\sigma_{y|xy}^2|{G}_{yy}(\bar{f})|^2}, 
\label{GAfreq}
\end{equation}
which can be written also as $a_{Y}(\bar{f})=A_Y+\bar{a}_{Y}(\bar{f})$, showing that it consists of a frequency-independent part equal to the time-domain GA (\ref{GAtime}) and of a frequency-specific part quantified by (\ref{GAbarfreq}). 
Remarkably, the spectral GA measure (\ref{GAfreq}) is zero over all frequencies in the absence of internal dynamics in the target process, i.e. $a_{Y}(\bar{f})=0$ $\forall \bar{f}$ if $a_{yy,k}=0$ $\forall k$,
and satisfies the spectral integration property, i.e.
\begin{equation}
    A_{Y} = 2 \int_0^{\frac{1}{2}} a_{Y}(\bar{f})\,d\bar{f}
\label{GAtime_freq}.
\end{equation}
In the next section, the properties of the new GA measures defined here will be investigated in simulated coupled processes.

\section{Theoretical examples}
In this section, we study the behavior of the measures of GC, GI and GA presented above using simulated AR processes. First, we simulate open-loop (\textit{Sect. III.A}) and closed-loop (\textit{Sect. III.B}) bivariate AR processes where the exact profiles of the spectral measures are computed (with sampling frequency $f_s=1$ Hz) from the true values imposed for the AR parameters.
Then, we consider a multivariate system where the dynamics of two interacting processes are perturbed by a third process which is not modelled in the calculation of GC, GI and GA (\textit{Sect. III.C}); in this case, estimations are performed from finite-length realizations of the three processes.
Finally, in \textit{Sect. III.D} we discuss the results of the simulations, using them to support the comparison and interpretation of the time-domain and spectral measures of GC, GI and GA.

\subsection{Open-Loop System}
The first simulation reproduces a bivariate AR process where the driver $X$ and the target $Y$ exhibit autonomous oscillations at different frequencies, and where a causal interaction from $X$ to $Y$ is simulated. The process is defined as: 
\begin{equation}
\begin{aligned}
X_n&=a_{x,1}X_{n-1}+a_{x,2}X_{n-2}+U_n \\
Y_n&=a_{y,1}Y_{n-1}+a_{y,2}Y_{n-2}-c X_{n-1}+V_n
\label{sim_open_loop}
\end{aligned}
\end{equation}
where $U$ and $V$ are independent Gaussian white noises with zero mean and unit variance. The autonomous oscillations in the two processes are obtained placing a pair of complex-conjugate poles, with modulus $\rho$ and phase $2\pi{f}$, in the complex plane representation of each process; the AR coefficients resulting from this setting are $a_1=2\rho \cos(2\pi f)$ and $a_2=-\rho^2$ \cite{faes2015information}. Here, we set $\rho_x=0.9, f_x=0.3$ Hz, so that the autonomous dynamics of $X$ are determined by the fixed coefficients $a_{x,1}= -0.556, a_{x,2}= –0.81$, and $\rho_y=b \cdot 0.8, f_y=0.1$ Hz, so that the strength of the autonomous dynamics of $Y$, which are determined by the coefficients $a_{y,1}, a_{y,2}$, depends on the parameter $b$.
Moreover, causal interactions are set from $X$ to $Y$ at lag $k=1$, with strength modulated by the parameter $c$.

\begin{figure} [h!]
    \centering
    \includegraphics[scale=0.45]{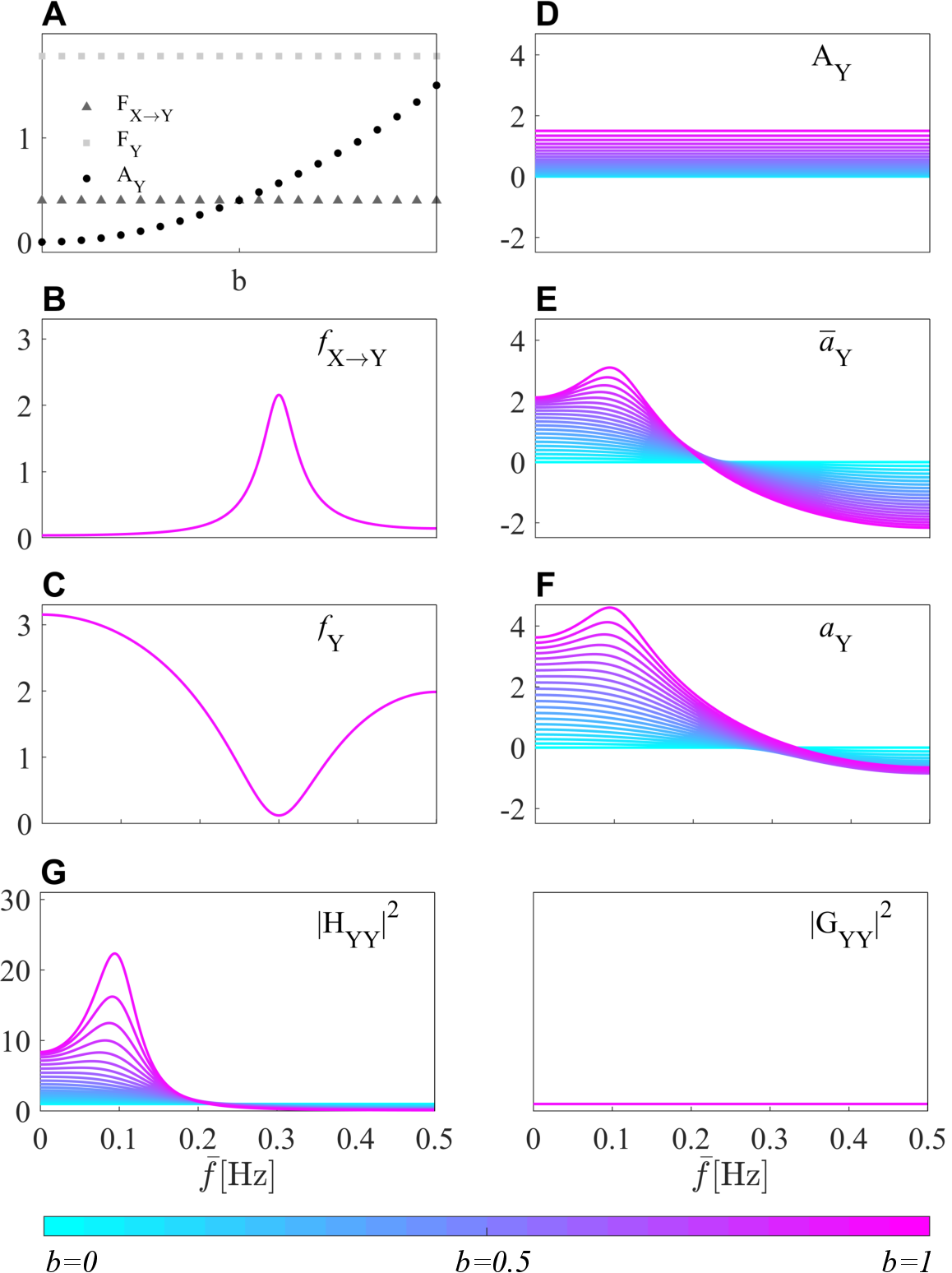}
    \caption{Dependence of the measures of Granger Causality, Isolation and Autonomy on the strength of the internal dynamics in the target process, modulated by the parameter $b$ in the open-loop system (\ref{sim_open_loop}). Plots depict: the time-domain values of the GC, GI, and GA measures $F_{X\to{Y}}, F_Y, A_Y$ (\textbf{A}); the spectral profiles of the GC measure $f_{X\to{Y}}(\bar{f})$ (\textbf{B}), GI measure $f_{Y}(\bar{f})$ (\textbf{C}) and GA measures $A_Y$, $\bar{a}_Y(\bar{f})$, $a_Y(\bar{f})$ (\textbf{D-F}); the spectral profiles of the transfer functions of the full (\ref{Y_ARX}) and restricted (\ref{Y_X}) models, $H_{yy}(\bar{f})$ and $G_{yy}(\bar{f})$ (\textbf{G}).}
    \label{sim_open_loop_1}
\end{figure}
\begin{figure} [h!]
    \centering
    \includegraphics[scale=0.45]{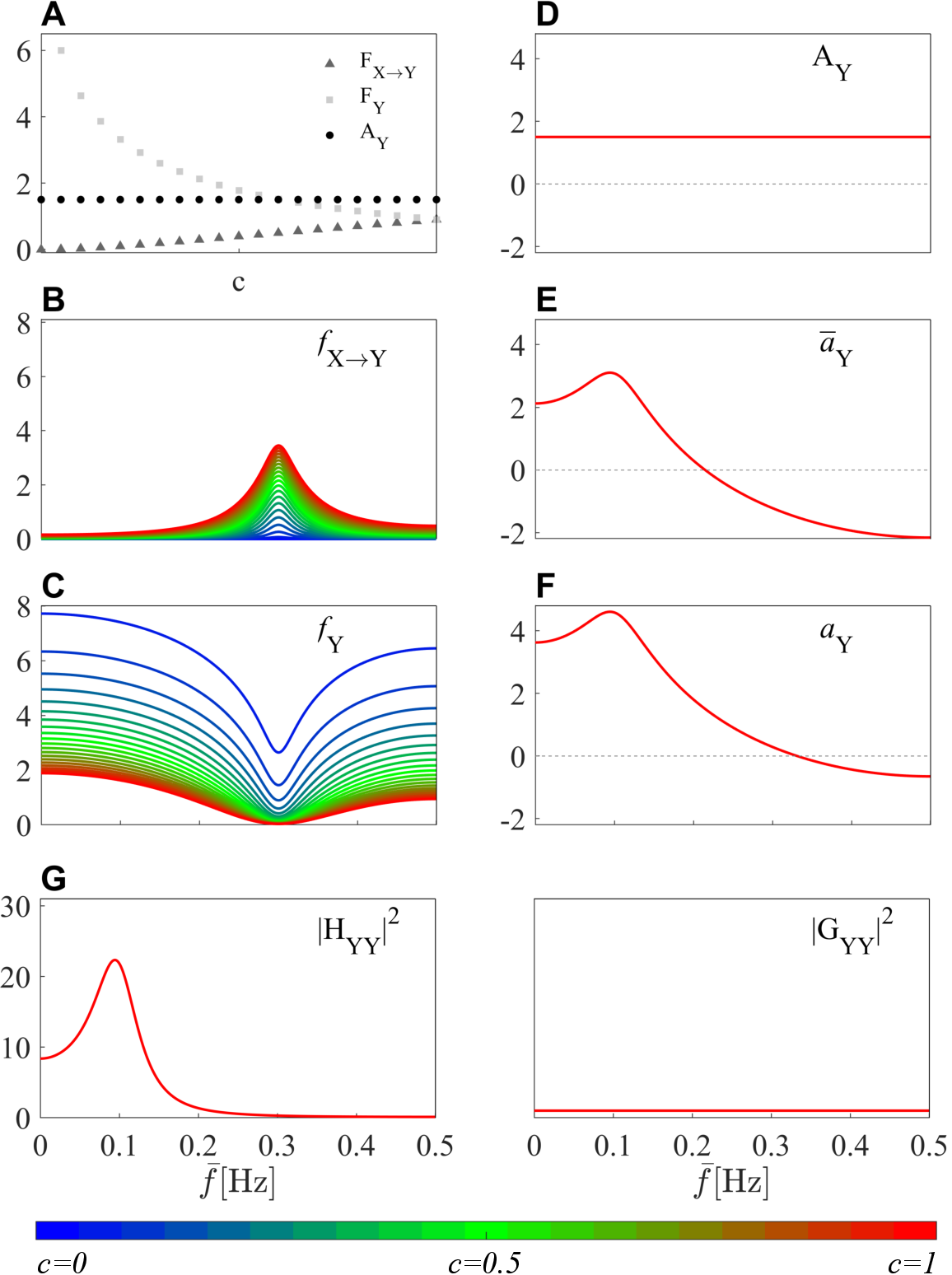}
    \caption{Dependence of the measures of Granger Causality, Isolation and Autonomy on the strength of the causal interaction from driver to target, modulated by the parameter $c$ in the open-loop system (\ref{sim_open_loop}). Plots depict: the time-domain values of the GC, GI, and GA measures $F_{X\to{Y}}, F_Y, A_Y$ (\textbf{A}); the spectral profiles of the GC measure $f_{X\to{Y}}(\bar{f})$ (\textbf{B}), GI measure $f_{Y}(\bar{f})$ (\textbf{C}) and GA measures $A_Y$, $\bar{a}_Y(\bar{f})$, $a_Y(\bar{f})$ (\textbf{D-F}); the spectral profiles of the transfer functions of the full (\ref{Y_ARX}) and restricted (\ref{Y_X}) models, $H_{yy}(\bar{f})$ and $G_{yy}(\bar{f})$ (\textbf{G}).}
    \label{sim_open_loop_2}
\end{figure}

We consider the two following settings: (i) progressive strengthening of the internal dynamics in the process $Y$ with stable causal interaction from $X$ to $Y$, obtained varying $b$ from 0 to 1 with fixed $c=0.5$; (ii) progressive strengthening of the causal interaction from $X$ to $Y$ with stable internal dynamics of $Y$, obtained varying $c$ from 0 to 1 with fixed $b=1$.
The time-domain values and spectral profiles of the measures of GC, GI and GA resulting from the two simulations are reported in Fig. \ref{sim_open_loop_1} and Fig. \ref{sim_open_loop_2}, respectively.
The GA measure $A_{Y}$ reflects exclusively the presence and strength of the autonomous dynamics in the target process $Y$, as it is null when $b=0$ and rises proportionally to $b$ in the first setting and is constant at varying the coupling from $X$ to $Y$ in the second setting (Fig. \ref{sim_open_loop_1}A and \ref{sim_open_loop_2}A, circles).
Analogously, the GC measure $F_{X\to{Y}}$ reflects exclusively the presence and strength of the causal coupling from $X$ to $Y$, being constant in case of fixed coupling $c=0.5$ (Fig. \ref{sim_open_loop_1}A, triangles) and increasing with $c$ when $b$ is kept constant (Fig. \ref{sim_open_loop_2}A, triangles).
The GI measure $F_Y$ is also affected only by the causal coupling (it is constant when $b$ varies with $c=0.5$, Fig. \ref{sim_open_loop_1}A, squares), and is complementary to the GC measure, as it exhibits an opposite trend (it decreases at increasing $c$, Fig. \ref{sim_open_loop_2}A, squares). 

The spectral measures of GC, GI and GA localize within specific frequency bands, related to the oscillations of the two processes, the effects described in the time domain. Indeed, the GC and GI measures $f_{X\to Y}(\bar{f})$ and $f_{Y}(\bar{f})$ exhibit respectively a peak and a valley at the frequency of the oscillation of $X$ that is transmitted to $Y$ (i.e., $f=0.3$ Hz, Fig. \ref{sim_open_loop_1}B,C). The shape of the spectral profile is modulated in both functions by the coupling parameter $c$ (Fig. \ref{sim_open_loop_2}B,C): when $c=0$ the GC is null at all frequencies and the GI takes the highest values; when $c$ rises towards 1 the GC shows a more and more prominent peak at $0.3$ Hz while the GI flattens progressively.
As regards the GA measure $a_Y(\bar{f})$, reported for the two simulations in Fig. \ref{sim_open_loop_1}F and Fig. \ref{sim_open_loop_2}F, we show its decomposition into a constant part equal to the time-domain GA measure $A_Y$ (Figs. \ref{sim_open_loop_1}D,\ref{sim_open_loop_2}D) and a variable part $\bar{a}_Y(\bar{f})$ whose frequency average is zero (Figs. \ref{sim_open_loop_1}E,\ref{sim_open_loop_2}E). The spectral GA shows its highest values at the frequency of the autonomous oscillations imposed in the target process (i.e., $f=0.1$ Hz, Fig. \ref{sim_open_loop_2}F); when the parameter determining the strength of this oscillation increases from $b=0$ to $b=1$, the spectral GA measures varies from a flat null profile up to a shape with a well-defined peak (Fig. \ref{sim_open_loop_1}F).

Fig. \ref{sim_open_loop_1}G and Fig. \ref{sim_open_loop_2}G report the spectral profiles of the transfer functions of the full (\ref{Y_ARX}) and restricted (\ref{Y_X}) models, $|H_{yy}(\bar{f})|^2$ and $|G_{yy}(\bar{f})|^2$ respectively, obtained varying the parameters $b$ and $c$.
These profiles illustrate how the predictable autonomous dynamics of the target process, in this simulation located at $0.1$ Hz, are captured in the full model by the transfer function $H_{yy}(\bar{f})$, but not in the reduced model by $G_{yy}(\bar{f})$, which indeed is flat. This corroborates the choice of the ratio between $|H_{yy}(\bar{f})|^2$ and $|G_{yy}(\bar{f})|^2$ as a meaningful index $\bar{a}_Y(\bar{f})$ displaying a peak at the frequency of the target autonomous dynamics (Figs. \ref{sim_open_loop_1}E, \ref{sim_open_loop_2}E).

\subsection{Closed-Loop System}
The second simulation reproduces a bivariate AR process where the driver $X$ and the target $Y$ exhibit autonomous oscillations at different frequencies and reciprocally interact in a closed loop. The process is defined as: 
\begin{equation}
\begin{aligned}
X_n&=a_{x,1}X_{n-1}+a_{x,2}X_{n-2}-d Y_{n-1}+U_n \\
Y_n&=a_{y,1}Y_{n-1}+a_{y,2}Y_{n-2}-c X_{n-1}+V_n
\label{sim_closed_loop}
\end{aligned}
\end{equation}
where $U$ and $V$ are independent Gaussian white noises with zero mean and unit variance.
The parameters $a_{\cdot,\cdot}$ were set as in the first simulation (\textit{Sect. III.A}) to obtain autonomous oscillations at 0.3 Hz for $X$ and at 0.1 Hz for $Y$. In addition to the causal interaction from $X$ to $Y$ modulated by $c$, a causal interaction is set from $Y$ to $X$ at lag $k=1$, with strength modulated by the parameter $d$.

\begin{figure} [h!]
    \centering
    \includegraphics[scale=0.415]{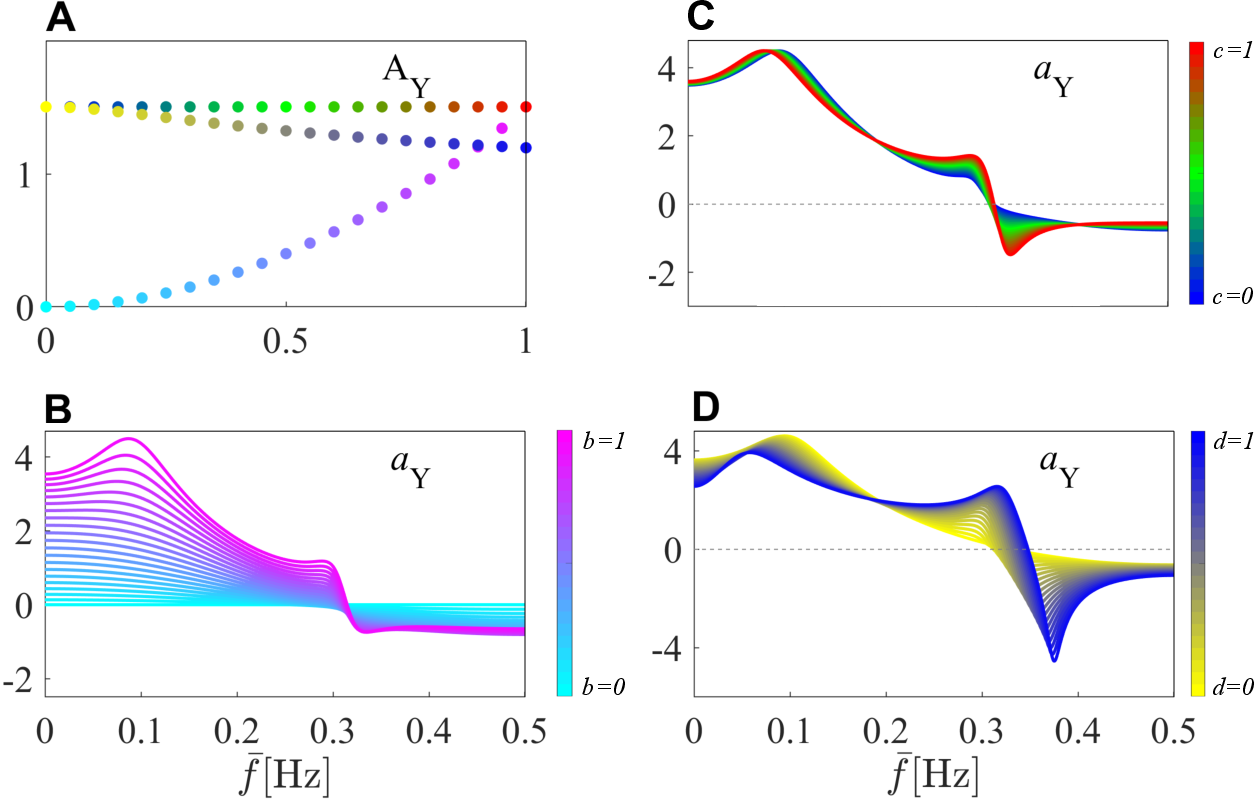}
    \caption{Dependence of the measure of Granger Autonomy on the parameters $b$, $c$ and $d$ of the closed-loop system (\ref{sim_closed_loop}), modulating respectively the strength of the internal dynamics in the target process, the causal coupling from $X$ to $Y$ and the causal coupling from $Y$ to $X$.
    Plots depict the time-domain values of the GA measure $A_Y$ at varying $b$ (light blue to magenta circles), $c$ (blue to red circles) or $d$ (yellow to blue circles) in the range $[0,1]$ (\textbf{A}), and the spectral profiles of the GA measure $a_Y(\bar{f})$ obtained varying $b$ (\textbf{B}), $c$ (\textbf{C}), and $d$ (\textbf{D}).}
    \label{sim_closed_loop_all}
\end{figure}

We consider the three following settings: (i) progressive strengthening of the internal dynamics in the process $Y$ with stable causal interactions, obtained varying $b$ from 0 to 1 with fixed $c=0.5$ and $d=0.2$; (ii) progressive strengthening of the causal interaction from $X$ to $Y$ with stable internal dynamics of $Y$ and causal interaction from $Y$ to $X$, obtained varying $c$ from 0 to 1 with fixed $b=1$ and $d=0.2$; (iii) progressive strengthening of the causal interaction from $Y$ to $X$ with stable internal dynamics of $Y$ and causal interaction from $X$ to $Y$, obtained varying $d$ from 0 to 1 with fixed $b=1$ and $c=0.5$.
Fig. \ref{sim_closed_loop_all} reports the time-domain values and the frequency-domain profiles of the GA obtained in the three settings. For the sake of brevity, the trends of the GC and GI measures are not reported because they are identical to those of the first simulation (Figs. \ref{sim_open_loop_1}, \ref{sim_open_loop_2}) despite the addition of the causal interaction from $Y$ to $X$.

Fig. \ref{sim_closed_loop_all}A shows that the time-domain GA increases with the strength of the internal dynamics modulated by $b$ (circles, light blue to magenta), and remains constant increasing the strength of the interaction $X \rightarrow Y$ modulated by $c$ (circles, blue to red).
On the other hand, stronger interactions $Y \rightarrow X$ obtained increasing $d$ determine a slight decrease of the time-domain GA $A_Y$ (circles, yellow to blue).
The spectral expansion of the GA measure allows to identify the frequency bands where the internal dynamics are localized. Indeed, Figs. \ref{sim_closed_loop_all}B,C,D reveal that the spectral profile of the GA measure exhibits a peak at the frequency of the autonomous oscillation imposed in the target process (i.e., $0.1$ Hz). This peak is clearly modulated in amplitude by the strength of the internal dynamics of $Y$ (parameter $b$, Fig. \ref{sim_closed_loop_all}B), while it changes only slightly at varying the causal interactions between $X$ and $Y$ (parameters $c$ and $d$), showing small amplitude and frequency modulations (Fig. \ref{sim_closed_loop_all}C,D).
Furthermore, the imposition of a feedback effect from $Y$ to $X$ determines a modification of the spectral profile of $a_Y(\bar{f})$, with the emergence of a second peak around the frequency of the autonomous oscillations of $X$ ($\sim 0.3$ Hz) and of a a reverse peak at higher frequencies (Fig. \ref{sim_closed_loop_all}B,C,D). The spectral integration property allows to ascribe the decrease of the time-domain GA for high values of $d$ to this behavior, as the negative peak prevails over the positive one at high frequencies, while the low-frequency peak shows preserved or slightly larger amplitude.

\subsection{System with Unobserved Confounders}
In this section, we study the behavior of the measures of GC, GI and GA, computed as described in \textit{Sect. II} for two processes $X$ and $Y$, when their dynamics are affected by an ``unobserved" process $Z$. To do this, we simulate the three-variate process defined as
\begin{equation}
\begin{aligned}
&X_n=a_{x,1}X_{n-1}+a_{x,2}X_{n-2}+U_n,\\
&Y_n= a_{y,1}Y_{n-1} + a_{y,2}Y_{n-2} -0.8 X_{n-1} -a Z_{n-1} + V_n \\
&Z_n=a_{z,1}Z_{n-1}+a_{z,2}Z_{n-2}+W_n,
\end{aligned}
\label{XYZsimu}
\end{equation}
where $U$, $V$ and $W$ are independent Gaussian white noises with zero mean and unit variance.
The coefficients $a_{\cdot,\cdot}$ are set to obtain autonomous oscillations in the processes depending on the modulus $\rho$ and phase $2\pi{f}$ of three pairs of complex-conjugate poles. Here, we set $\rho_x=0.9, f_x=0.3$ Hz, $\rho_y=b \cdot 0.8, f_y=0.1$ Hz, and $\rho_z=0.8, f_z=0.2$ Hz; the strength of the autonomous dynamics of $Y$ depends on the parameter $b$. Moreover, causal interactions are set at lag 1 both from $X$ to $Y$, with fixed strength 0.8, and from $Z$ to $Y$, with strength weighed by the parameter $a$.

The analysis is performed on realizations of the three processes generated by feeding (\ref{XYZsimu}) with white noise observations, and then computing the spectral GC, GI and GA measures on the time series relevant to the processes $X$ and $Y$.
We consider three parameter settings:
(\textit{I}) $a=0, b=0$, to simulate the absence of autonomous dynamics in the target process $Y$ and of effects from the unobserved process $Z$;
(\textit{II}) $a=0, b=0.8$, to simulate the presence of autonomous dynamics in $Y$ without effects from $Z$;
(\textit{III}) $a=0.8, b=0$, to simulate the effect of the unobserved process $Z$ on the target $Y$ in the absence of autonomous dynamics.
For each setting, 100 realizations of (\ref{XYZsimu}) were generated, each of length $N=500$ points, and the spectral measures of GC, GI and GA were estimated after identifying the bivariate AR model fitting the time series of $X$ and $Y$; the model order was set using the Akaike Information Criterion (AIC) \cite{faes2012measuring}.

\begin{figure}[htbp]
\centerline{\includegraphics[scale=0.375]{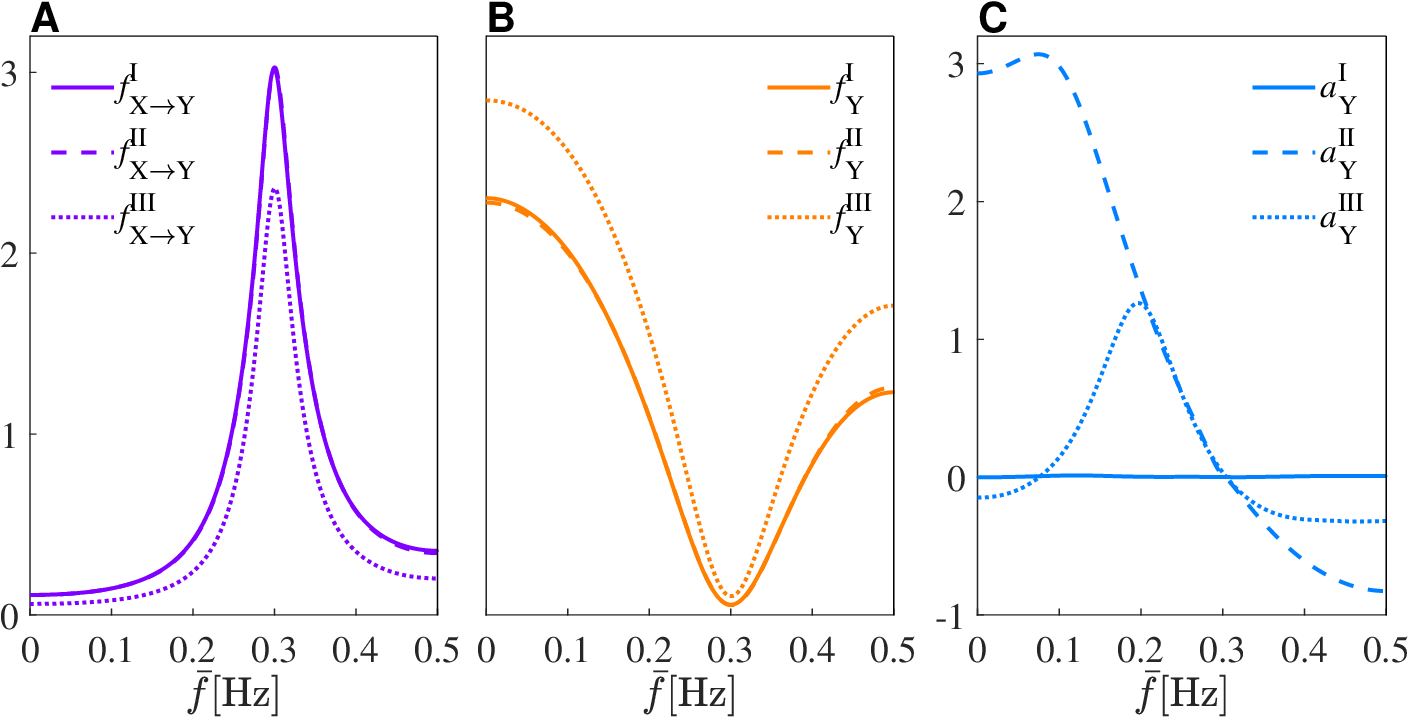}}
\caption{Analysis of Granger Causality, Isolation and Autonomy for the simulated system with unidirectional interactions and confounding effects.
Plots depict the spectral profiles, expressed as average over 100 runs of (\ref{XYZsimu}), of the measures of GC ($f_{X\rightarrow{Y}}(\bar{f})$, \textbf{A}), GI ($f_{Y}(\bar{f})$, \textbf{B}) and GA ($a_{Y}(\bar{f})$, \textbf{C}) computed in the absence of autonomous dynamics of $Y$ and confounding effects from $Z$ to $Y$ ($a=0$, $b=0$, continuous lines), in the presence of autonomous dynamics only ($a=0$, $b=0.8$, dashed lines), and in the presence of confounding effects only ($a=0.8$, $b=0$, dotted lines).}
\label{unobserved_driver}
\end{figure}


The results of the analysis are reported in Fig. \ref{unobserved_driver}, showing the average spectral profiles of the GC, GI and GA measures in the three simulation conditions. The profiles of GC and GI are very similar in the three cases, revealing a clear peak of the GC, and a corresponding minimum of the GI, at the frequency of the causal interaction imposed from $X$ to $Y$ ($f_x=0.3$ Hz, Fig. \ref{unobserved_driver}A,B). This documents that the presence of the unobserved confounder $Z$ acting only on the analyzed target $Y$ does not alter significantly the causal interactions from $X$ to $Y$.
On the other hand, the profiles of GA are substantially different in the three cases: Fig. \ref{unobserved_driver}C shows that the GA stays uniformly at the zero level when both autonomous target dynamics and confounding effects are absent (continuous line), peaks at $\sim 0.1$ Hz when only the autonomous dynamics are present (dashed line), and peaks at $\sim 0.2$ Hz when only the confounding effects are present (dotted line). This documents that the proposed spectral measure of GA captures not only the autonomous dynamics of $Y$, but also the regular dynamics simulated in $Z$ and transmitted to $Y$ via the parameter $a$.

\subsection{Interpretation and Comparison of GC, GI and GA}

The reported simulations depict the theoretical properties of the measures of GC, GI and GA developed in this work.
We have shown that, in a bivariate process $\{X,Y\}$, the GC and GA measures capture selectively the causal interaction from $X$ to $Y$ and the autonomous dynamics of $Y$, respectively, either globally or at specific frequencies when measured in the time or frequency domains. The GI measure behaves in a complementary way to the GC, decreasing with the strength of the causal interactions and thus reflecting the degree of isolation of $Y$.
Importantly, the time- and frequency-domain formulations of GC, GI and GA are strictly connected by the spectral integration property, and the spectral representation allows to identify the oscillations for which causal, non-causal and autonomous effects take place.
This property can be useful to detect variations in the strength of effects which are confined within specific frequency bands and can be missed if investigated in the time domain only.

However, while these interpretations emerge strikingly in a bivariate system with unidirectional coupling, they can be challenged when more complex dynamics arise in the presence of closed-loop or multivariate interactions.
For instance, in the second simulation (\textit{Sect. III.B}) we showed that the GA measure is influenced by the imposition of a feedback effect from the target to the driver; 
a similar behavior was documented in a previous work by the conditional self entropy measure \cite{faes2015information}, which is formally equivalent to the time-domain GA.
Herein, this dependence is localized in frequency via the proposed spectral GA measure, which exhibits an irregular profile with the appearance of a positive peak and a reverse one around the frequency of the autonomous oscillation in the driver system (see, e.g., Fig \ref{sim_closed_loop_all}B,C,D).
Furthermore, the third simulation (\textit{Sect. III.C}) showed that, when an unobserved process has effects on the target, these effects may alter the spectral profile of the GA measure in a way similar to that of autonomous dynamics (see Fig. \ref{unobserved_driver}C).
This may have implications in practical applications, e.g. when multiple physiological systems interact but only two of them are monitored in a bivariate analysis. 

Further insights on the concepts of “autonomy" and “isolation" are provided in the supplemental material of this paper (\textit{Sect. S2}), where their meanings are discussed in terms of target predictable dynamics and non-causal spectral power and the relevant measures of GA and GI are compared in theoretical examples. These examples highlight the distinct nature of the two measures, as they evidence how autonomy and isolation can coexist in the same bivariate process, be selectively present, or be both absent.

As regards the relation between the concepts of isolation and causality, quantified respectively by the GI and GC measures, we evidence that they are clearly complementary, as an increase in the causal part of the spectrum implies a decrease of the isolated part and vice versa (see eqs. (\ref{spectraldecomposition}), (\ref{GewekeGCfreq}) and (\ref{nGCfreq})).
However, differently from the corresponding non-logarithmic DC measures (\ref{DC}) and (\ref{DC_diag}) which sum to $1$ at each frequency, the relation between the GC and GI measures is not trivial. In fact, the logarithmic transformation, which provides information-theoretic meanings to GC and GI, makes their sum to vary across frequencies, and this aspect may differentiate their behavior in practical computations; we show an example in \textit{Sect. IV}.


\section{Application to cerebrovascular data}

This section reports the practical computation of the spectral measures of GC, GI and GA on cerebrovascular time series measured in healthy controls and subjects prone to develop postural-related syncope \cite{bari2016nonlinear,faes2013investigating}. 
Interactions between mean arterial pressure (MAP) and mean cerebral blood flow velocity (MCBFV) time series have been largely studied to investigate the cerebrovascular (CB) control and dynamic cerebral autoregulation (CA) in a variety of physiopathological conditions \cite{aaslid1989cerebral,paulson1990cerebral,faes2013investigating,bari2017cerebrovascular}.
CB interactions are largely determined by the so-called pressure-to-flow link, according to which variations of MAP drive similar changes in CBFV but also trigger CA responses whereby an homeostatic regulation of CBFV is looked for \cite{aaslid1989cerebral,paulson1990cerebral}.
Here, we hypothesize that spectral indexes quantifying both the causal effects of MAP on MCBFV and the autonomous dynamics of MCBFV can identify better than the more commonly used time-domain indexes the alteration of the physiological control mechanisms related to CB interactions and to CA occurring with postural stress in subjects with poor orthostatic tolerance.

\begin{table*}[htbp]
\captionsetup{font=footnotesize, format=hang, labelsep=colon, justification=raggedright, skip=3pt}
\caption{Time-domain indexes of mean and variance of MAP ($\mu_{X}$ $[mmHg]$, $\sigma^2_{X}$ $[mmHg^2]$) and MCBFV ($\mu_{Y}$ $[cm/s]$, $\sigma^2_{Y}$ $[(cm/s)^2]$) shown as mean$\pm$std.dev. across subjects for the different groups (non-SYNC, SYNC) and experimental conditions (REST, ET, LT). Statistically significant differences assessed via paired Wilcoxon test with Bonferroni-Holm correction for multiple comparison: $^*$, REST vs. ET, REST vs. LT; $^{\#}$, ET vs. LT.}
\begin{center}
\begin{tabular}{ c|c|c|c|c|c|c } 
\hline\hline
  & \multicolumn{3}{c|}{\textbf{non-SYNC}} & \multicolumn{3}{c}{\textbf{SYNC}} \\
  \hline
  & \textbf{REST} & \textbf{ET} & \textbf{LT} & \textbf{REST} & \textbf{ET} & \textbf{LT} \\ 
  \hline
  \textbf{$\mu_{X}[mmHg]$} & \scriptsize{$98.84\pm17.33$} & \scriptsize{$95.16\pm12.17$} & \scriptsize{$92.94\pm11.61$} & \scriptsize{$84.42\pm13.96$} & \scriptsize{$97.16\pm17.50^*$} & \scriptsize{$93.53\pm15.64^*$} \\
 \hline
 \textbf{{$\sigma_{X}^2[mmHg^2]$}} & \scriptsize{$14.22\pm14.79$} & \scriptsize{$15.38\pm9.11$} & \scriptsize{$14.56\pm9.46$} & \scriptsize{$9.08\pm6.78$} & \scriptsize{$13.61\pm7.26$} & \scriptsize{$15.08\pm6.99$} \\
 \hline
 \textbf{$\mu_{Y}[cm/s]$} & \scriptsize{$72.02\pm23.14$} & \scriptsize{$62.12\pm21.52^*$} & \scriptsize{$61.09\pm15.72^*$} & \scriptsize{$64.42\pm17.25$} & \scriptsize{$56.25\pm17.06^*$} & \scriptsize{$48.12\pm18.08^{*\#}$} \\
 \hline
  \textbf{$\sigma_{Y}^2[(cm/s)^2]$} & \scriptsize{$12.74\pm8.20$} & \scriptsize{$20.42\pm11.42^*$} & \scriptsize{$15.42\pm10.38$} & \scriptsize{$34.67\pm72.69$} & \scriptsize{$41.56\pm95.99$} & \scriptsize{$32.20\pm56.31$} \\
 \hline\hline
\end{tabular}
\end{center}
\label{tab1}
\end{table*}

\subsection{Subjects and Experimental Protocol}
The analyzed time series belong to a database previously collected to study the short-term physiological regulation in subjects prone to neurally-mediated syncope and healthy controls via the analysis of spontaneous variability of systemic variables \cite{faes2013investigating, bari2016nonlinear}. The study included 13 subjects (age: $28\pm{9}$ years; 5 males) with previous history of unexplained syncope (SYNC, reporting $>$3 syncope events in the previous 2 years) and 13 age-matched healthy subjects (non-SYNC, age: $27\pm{8}$ years; 5 males), enrolled at the Neurology Division of Sacro Cuore Hospital, Negrar, Italy. The protocol consisted of 10 minutes of recording in the resting supine position, followed by 60° head-up tilt test. All SYNC subjects experienced presyncope signs (i.e., a vasovagal episode characterized by hypotension and reflex bradycardia leading to partial loss of consciousness) during the tilt session; when signs were reported, the subject was returned to the resting position and a spontaneous recovery occurred. None of the non-SYNC subjects experienced presyncope symptoms during tilt. 

The signals analyzed in this study are the electrocardiogram (ECG, lead II), the continuous AP measured at the level of middle finger through a photopletysmographic device (Finapres, Enschede, The Netherlands), and the CBFV signal measured at the level of the middle cerebral artery by means of a transcranial doppler (TCD) ultrasonographic device (Multi-Dop T, Compumedics, San Juan Capistrano, CA, USA). 
From these signals, the variability series of MAP and MCBFV were extracted on a beat-to-beat basis by taking the average of the AP and CBFV signals measured between the local minima occurring in the signals after each heartbeat detected from the ECG. For each subject, three sequences of 250 consecutive synchronous values of MAP and MCBFV were selected for the analysis, corresponding to the following experimental conditions: (i) supine rest (REST); (ii) early tilt (ET), starting after the onset of the head-up tilt maneuver, excluding physiological transients and limiting the influence of non-stationarities over the analysis; (iii) late tilt (LT), starting at least 5 minutes after the onset of the tilt maneuver for non-SYNC subjects, and occurring just before the pressure decrease due to presyncope
for SYNC subjects.
Selection of the sequences was performed randomly in each experimental condition and repeated if non-stationarities of the mean and the variance were present. The series were visually inspected and eventually corrected through cubic spline interpolation, with corrections not exceeding the 5\% of the overall length of the sequence. 

Further information about the experimental protocol, signal acquisition and variability series extraction can be found in \cite{faes2013investigating, bari2016nonlinear}.

\subsection{Data Analysis}
The time series extracted for each subject in the three experimental conditions were regarded as realizations of the MAP (process $X$) and MCBFV (process $Y$) discrete-time processes. These processes were assumed as uniformly sampled with a sampling frequency equal to the inverse of the mean heart period $<HP>$ ($f_s=\frac{1}{<HP>}$).

First, classical time domain markers such as the mean and variance of MAP ($\mu_X$, $\sigma_{X}^2$) and MCBFV ($\mu_Y$, $\sigma_{Y}^2$) were computed.
Then, the series were pre-processed reducing the slow trends with an AR high-pass filter (zero phase; cut-off frequency 0.0156 Hz) and removing the mean value. A bivariate AR model in the form of (\ref{2ARmodel}) was fitted on each pair of pre-processed series using vector least-squares identification and setting the model order $p$ according to the multivariate version of the AIC (maximum scanned order $=14$); the series and the PSD profiles were visually inspected and model orders were manually fixed where necessary, i.e. where too many or few spectral peaks were observed.
After AR identification, the time-domain and spectral measures of GC, GI and GA were obtained computing the parameters of the restricted models (\ref{ARmodel}) and (\ref{Y_X}) from the estimated full-model parameters and then applying the derivations presented in \textit{Sect. II}. 
Fig. \ref{repr_series} reports an example of MAP and CBFV time series, together with their estimated PSDs and spectral GC, GI and GA profiles, measured for a representative subject.
The spectral measures of GC from MAP to MCBFV, GI of MCBFV, and GA of MCBFV (respectively, $f_{X\to{Y}}$, $f_{Y}$, and $a_{Y}$) were integrated within the two frequency bands of physiological interest for CB variability, i.e. the very-low frequency (VLF, $f\in [0.02,0.07]$ Hz) and low frequency (LF, $f\in [0.07,0.2]$ Hz) \cite{claassen2016transfer}, as well as over the whole frequency range $[0,f_s/2]$ to get the time-domain values $F_{X\to{Y}}$, $F_{Y}$, and $A_Y$.


\subsubsection{Surrogate data analysis}
To test the statistical significance of the GC, GI and GA measures, a bootstrap method using explicit model equations extracted from the data \cite{schreiber2000surrogate} was implemented. The method generates surrogates of the observed time series $X$ and $Y$ according to the null hypothesis of absence of causal coupling from $X$ to $Y$ (H$_1$), or absence of internal dynamics within the process $Y$ (H$_2$).
Specifically, each original MAP series was fitted with the ARX model (\ref{X_ARX}), while the corresponding MCBFV series was fitted with the AR model (\ref{ARmodel}) to test H$_1$ and with the X model (\ref{Y_X}) to test H$_2$.
Then, in each case, pairs of surrogate time series were generated feeding the models with noise realizations obtained shuffling randomly the samples of the estimated residuals.

\begin{figure*}[htbp]
\centerline{\includegraphics[scale=0.45]{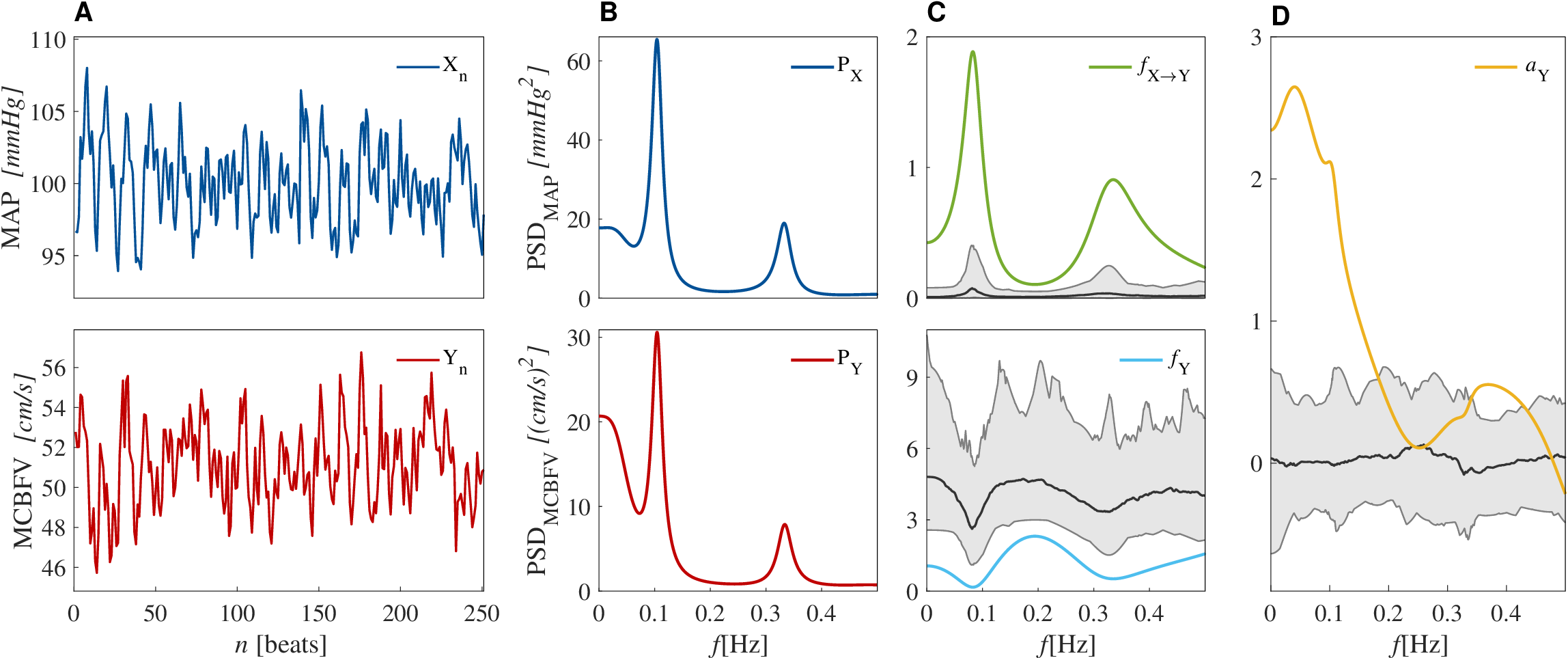}}
\caption{Example of Granger Causality, Isolation and Autonomy analyses for a representative non-SYNC subject in the REST condition (model order: $p=7$). (\textbf{A}) MAP  and CBFV time series measured as realizations of the processes $X$ and $Y$ for this subject. (\textbf{B}) PSD profiles of the MAP series, $P_X$, and of the MCBFV series, $P_Y$. (\textbf{C}) Spectral profiles of the GC from MAP to MCBFV ($f_{X\to{Y}}$, green) and of the GI of MCBFV ($f_Y$, light blue). (\textbf{D}) Spectral profile of the GA of MCBFV ($a_Y$, yellow). In \textbf{C} and \textbf{D}, the distributions of the spectral GC, GI and GA measures computed from surrogate time series are depicted as shaded areas, median (black lines) and percentiles (grey lines, computed with a $5\%$ significance level).}
\label{repr_series}
\end{figure*}

One-hundred pairs of surrogate time series were obtained iterating this procedure, and the time-domain and spectral measures of GC, GI and GA were computed at each iteration. The significance of the measures, computed either in the time domain or integrating the spectral function over the VLF or LF bands, was assessed comparing the values obtained on the original time series with the confidence limits of the surrogate distribution computed with 5{\%} significance. Specifically, the GC, GI and GA were deemed as statistically significant if their value was respectively above the 95$^{th}$ percentile of the GC distribution over surrogates generated under $H_1$, below the $5^{th}$ percentile of the GI distribution over surrogates generated under H$_1$, and above the $97.5^{th}$ or below the $2.5^{th}$ of the GA distribution over surrogates generated under H$_2$. A representative example is illustrated in Fig. \ref{repr_series}C,D. 

\subsubsection{Statistical analysis}
The distributions of the time-domain markers as well as of GC, GI and GA computed across subjects for each group (SYNC and non-SYNC) were tested for normality using the Anderson-Darling test. Since the hypothesis of normality was rejected for most distributions, and given the small sample size, non-parametric tests were employed to assess the statistical significance of the differences of each index across conditions.
Specifically, the one-way Friedman test was employed to assess the significance of the differences across conditions, followed in case of rejection by a post-hoc pairwise comparison carried out through the paired Wilcoxon test with Bonferroni-Holm correction for multiple comparison ($n=3$) to assess the differences between pairs of distributions (REST vs. ET, REST vs. LT, ET vs. LT). 
All the statistical tests were carried out with 5\% significance level.

Pre-processing of the time series, GC, GI, and GA measures estimation, surrogate data analysis and statistical analysis were all performed using MATLAB 2021b (The Mathworks, Inc.).

\begin{figure*}
\centerline{\includegraphics[width= \textwidth,keepaspectratio]{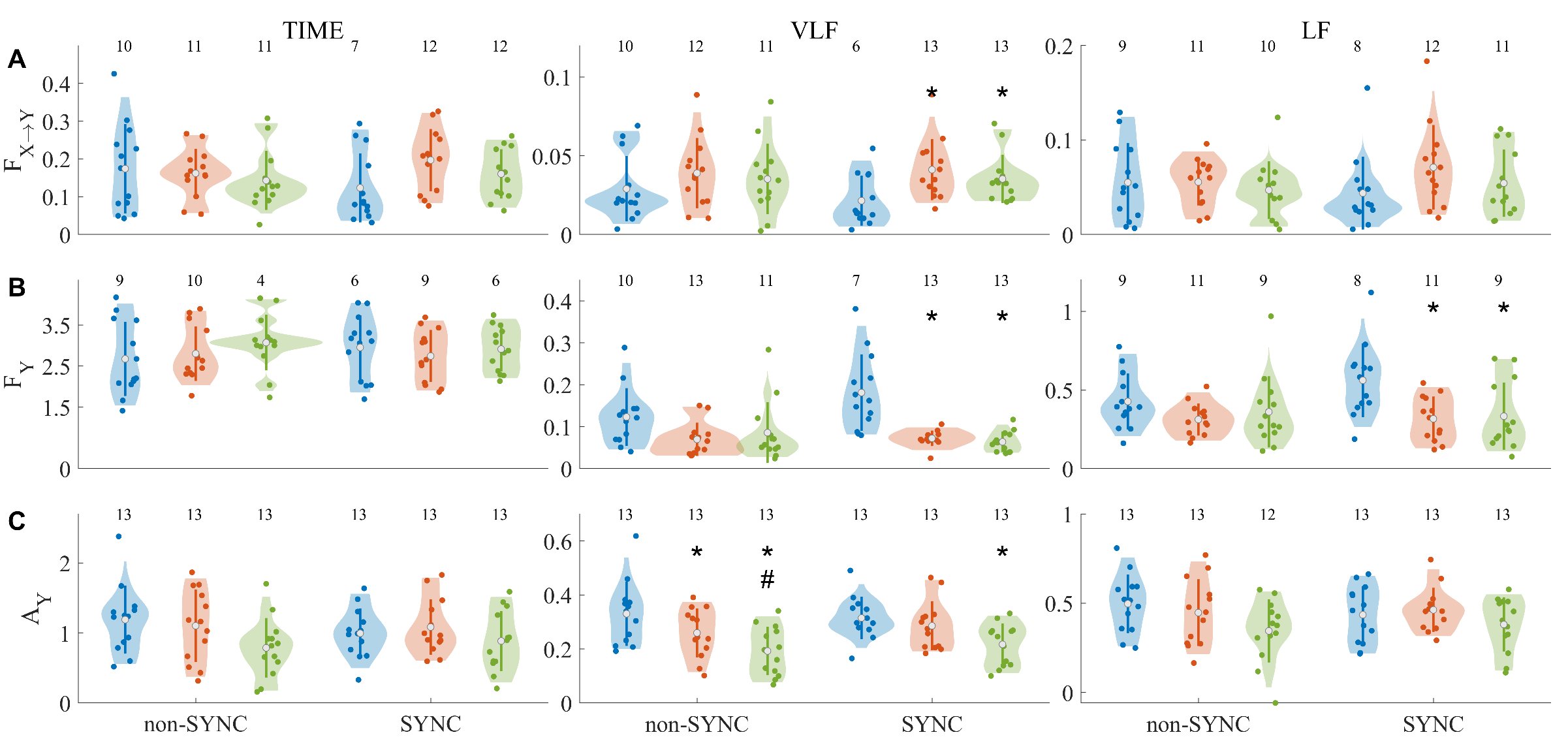}}
\caption{Analysis of Granger Causality, Isolation and Autonomy for the cerebrovascular time series measured in subjects prone to postural syncope (SYNC) and healthy controls (non-SYNC). Plots depict the distributions across subjects (individual values and violin-plots) of the GC from MAP to MCFV (\textbf{A}), of the GI of MCBFV (\textbf{B}), and of the GA of MCBFV (\textbf{C}) computed in the time domain (left plots) and integrating the spectral functions within the VLF band ($[0.02,0.07]$ Hz, middle plots) and the LF band ($[0.07,0.2]$ Hz, right plots). For each group and band, measures are computed at REST (blue) and during the early phase (ET, orange) and the late phase (LT, green) of head-up tilt; for each distribution, the mean and interquartile range are depicted by the white circle and vertical line, respectively, while the width of the violin plot denotes probability density.
Values above each distribution indicate the number of subjects for which the measure was deemed as significant according to surrogate data analysis.
Statistically significant differences assessed via paired Wilcoxon test with Bonferroni-Holm correction for multiple comparison: $^*$, REST vs. ET, REST vs. LT; $^{\#}$, ET vs. LT.}
\label{appl_cerebro}
\end{figure*}

\subsection{Results and Discussion}
Table \ref{tab1} depicts the results - in terms of time-domain markers (mean $\mu$ and variance $\sigma^2$) of the MAP and MCBFV series computed in the two analyzed groups during the three experimental conditions.
The trends of these markers document the expected CB response to the orthostatic stress in subjects prone to syncope and controls \cite{grubb1991cerebral, bari2017cerebrovascular, faes2013investigating}.
Specifically, the average MCBFV decreased significantly during tilt in both groups as a consequence of the physiologic cerebral vasoconstriction associated with the orthostatic challenge. In the SYNC group, the drop of $\mu_Y$ was more marked during LT and was accompanied by a significant increase, during both ET and LT compared to REST, of the average MAP, likely reflecting a progressive weakening of CA mechanisms which occurs with prolonged postural stress.
The variability of the two series did not show evident trends across conditions, except for an increase of $\sigma^2_X$ during ET. 


Fig. \ref{appl_cerebro} reports the results of the analysis of causal, isolated and autonomous dynamics performed in the time and frequency domains. All the time-domain measures
do not exhibit significant changes across the three analyzed experimental conditions  (Fig. \ref{appl_cerebro}, left plots). On the other hand, the evaluation of the same measures within the frequency bands of physiological interest for this application (i.e., VLF and LF) highlights some evident variations during the orthostatic stress, also differentiating the response between syncope subjects and healthy controls (Fig. \ref{appl_cerebro}, middle and right plots).
These different behaviors of time-domain and spectral measures evidence the need, for this physiological application, of assessing causal and autonomous dynamics in the frequency domain to capture mechanisms that remain otherwise hidden if a whole-band time-domain analysis is performed.

The spectral analysis reveals, for the SYNC group, a significant increase of the GC from MAP to MCBFV and a significant decrease of the GI of MCBFV during both epochs of head-up tilt (ET, LT) compared with REST (Fig. \ref{appl_cerebro}A,B); the changes are observed in the VLF band for both measures, where they occur together with a marked increase in the number of subjects for which the GC and GI were statistically significant according to the surrogate data analysis, and also in the LF band for the GI measure.
Methodologically, this finding confirms the simulation results showing that GC and GI provide complementary information, but also suggests that the two measures are not fully dependent on each other. Here, when assessed in specific frequency bands, the two measures describe physiological mechanisms with a different degree of discrimination: the tilt-induced enhancement of the influences of MAP on MCBFV is better captured by the GI measure.
Physiologically, the presence of stronger causal interactions along the pressure-to-flow link during tilt, detected in the subjects prone to develop postural syncope but not in the healthy controls, may be indicative of a defective CA, i.e. of a reduced intrinsic ability of the cerebral vascular bed to maintain a stable perfusion despite blood pressure changes.
Indeed, the increased causal coupling indicates that the variability of MCBFV is determined to a larger extent by the variability of MAP, and that the autoregulatory mechanism cannot respond fast enough to compensate for pressure changes. This interpretation agrees with that of previous studies in which a loss of CA has been associated with an increased link between AP and CBFV \cite{panerai1998frequency,zhang1998transfer}.
The physiological mechanisms leading to the weakening of CA in the subjects prone to syncope are complex, and possibly include hypercapnia (i.e., augmented arterial carbon dioxide pressure) \cite{panerai1999effect}, which has been associated to upwards shift of the coherence between MAP and CBFV at frequencies $<0.1$ Hz \cite{panerai1999effect},
and vasoconstriction (i.e., reduction in the diameter of large vessels) \cite{grubb1991cerebral}, which can have an effect on the measured MCBFV since the Doppler ultrasound measures blood flow velocity and not absolute flow.


As regards the autonomy measure, the spectral analysis evidences a progressive reduction of the GA of MCBFV computed in the VLF band moving from REST to ET, and from ET to LT (Fig. \ref{appl_cerebro}C); the decrease is evident and statistically significant for each pairwise comparison in the non-SYNC healthy controls, while it is less marked and significant only comparing REST vs. LT in the syncope subjects.
The decrease of GA with tilt indicates that the internal regulatory mechanisms of MCBFV acting in the VLF band loose progressively their strength during prolonged postural stress. As the decrease is evident particularly in the healthy controls, it seems to have a physiological rather than pathological origin; therefore, it should not regard the dynamic CA expressed in terms of interdependence between pressure and flow, which is indeed not efficiently represented by the GA measure. More likely, the decrease of GA reflects the reduced strength of exogenous effects, i.e., effects acting on MCBFV independently of MAP. Such effects might include the occurrence of hypocapnia with the orthostatic challenge in healthy subjects \cite{cencetti1997effect}, which may have an impact on arteriolar vessel caliber, and thus on blood flow velocity. This impact, which is not observed nor quantified in our AR model, might alter the autonomous dynamics of CBFV and thus enter the computation of GA. 
A simulation example investigating the effects of unobserved confounders is provided in \textit{Sect. III.C}.

This preliminary application presents some limitations.
First, since the small size of analyzed group of subjects may represent an issue when one aims to generalize results to an entire population, the use of larger datasets is needed to confirm the results obtained here.
Furthermore, in physiological applications where multiple complex interactions often arise, the effects of unobserved confounders are likely to occur as we have shown in the simulated settings. Therefore, the extension to multivariate datasets including signals possibly acting as confounders, as well as the development of multivariate extensions of the proposed measures of GC, GI and GA, are envisaged for future studies.

\section*{Conclusions}
The aim of this study was to explore, in addition to the well-known measure of Granger causality, the concepts of isolation and autonomy in coupled physiological processes, with emphasis on their frequency domain representation.
In this context, we develop a framework where already known and novel time-domain measures of GC, GI and GA are obtained as the full-frequency integral of their spectral counterparts. The framework allows quantification of the concepts of causality, isolation and autonomy either considering the overall dynamics of the observed bivariate process or the oscillations at specific frequencies of physiological interest.
Our theoretical derivations and experimental results document that the GI measure is complementary to GC but not trivially related to it, while GA reflects the regularity of the internal dynamics of the analyzed target process.

The frequency-domain formulation of GC, GI and GA is particularly useful for the analysis of dynamic processes which are rich of oscillatory content, as it allows to elicit physiological mechanisms which can be hidden in time domain due to the mixing with other spectral effects. This potential is demonstrated in our application to cerebrovascular interactions where the spectral measures highlight responses to postural stress which cannot be traced by the time-domain analysis.
In particular, our results suggest that GA quantifies the frequency-specific physiological response to postural stress of the slow CBFV oscillations, while GC and especially GI characterize the pathological response related to the impairment of the dynamic autoregulation of CBFV preceding the onset of postural-related syncope.

Possible limitations of the proposed approach stand in its bivariate formulation, which prevents from treating multivariate interactions among time series, and in the fact that the role of instantaneous effects among the processes, i.e. interactions occurring at zero lag \cite{chicharro2011spectral,pernice2022spectral}, is not explicitly addressed.
Therefore,
future developments of the proposed framework are envisaged to expand it toward the analysis of multivariate processes, so as to investigate the impact of unobserved confounders (such as, in our application, respiration and arterial carbon dioxide \cite{porta2008influence, panerai1999effect}) on the measures of causality, isolation and autonomy.
Methodologically, this can be done using vector driver processes in place of a single scalar process, leading
to quantify the degree of isolation of the target system in the multivariate case.
Moreover, the inclusion of instantaneous effects in the analyzed parametric models, though not always straightforward \cite{baccala2021frequency, nuzzi2021extending,pernice2022spectral}, is recommended to provide a complete picture of causal, isolated and autonomous effects emerging in the time and frequency domains from dynamic interactions.

\section*{Software Accessibility}
The Matlab Software relevant to this work is available for free download from \small{\url{https://github.com/LauraSparacino/GICA_toolbox}}.

\section*{Acknowledgment}
The Authors acknowledge Gianluca Rossato and Davide Tonon for providing the signals data processed in previous studies \cite{faes2013investigating, bari2016nonlinear} and re-analyzed in the present work.

\bibliographystyle{IEEEtran}

\end{document}


\title{A Method to Assess Granger Causality, Isolation and Autonomy in the Time and Frequency Domains: Theory and Application to Cerebrovascular Variability\\ \ \\Supplemental material}

\author{ 
\thanks{ }
\thanks{ }
\thanks{ }}

\maketitle


\section{Autoregressive models}

The classical autoregressive (AR) model description of a discrete-time, zero-mean stationary bivariate stochastic process $\textbf{S}_n=[X_{n} Y_{n}]^\intercal$ is given by \cite{chicharro2011spectral,faes2012measuring}
\begin{subequations}
\begin{align}
X_n&=\sum_{k=1}^p a_{xx,k}X_{n-k}+a_{xy,k}Y_{n-k}+U_{x|xy,n}, \label{X_ARX}\\
Y_n&=\sum_{k=1}^p a_{yx,k}X_{n-k}+a_{yy,k}Y_{n-k}+U_{y|xy,n}, \label{Y_ARX}
\end{align}
\label{2ARmodel}
\end{subequations}
where $p$ is the model order, defining the maximum lag used to quantify interactions, the coefficients $a$ quantify the time-lagged interactions within and between the two processes, and $U_{x|xy,n}$ and $U_{y|xy,n}$ are uncorrelated white noise processes with variance $\sigma_{x|xy}^2$ and $\sigma_{y|xy}^2$ collected in the 2x2 covariance matrix $\mathbf{\Sigma}$ ($\mathbf{\Sigma}=\begin{bmatrix}
\sigma_{x|xy}^2 & 0 \\
0 & \sigma_{y|xy}^2 \\
\end{bmatrix}$ if the bivariate process is strictly causal).
The model (\ref{2ARmodel}) is composed by two auto- and cross-regressive (ARX) models whereby each process is regressed both on its own past and on the past of the other process. In compact form, it can be formulated as $\textbf{S}_n=\sum_{k=1}^p \textbf{A}_{k}\textbf{S}_{n-k}+\textbf{U}_n$, where $\textbf{U}_n=[U_{x|xy,n} U_{y|xy,n}]^\intercal$ and $\textbf{A}_{k}$ is the 2x2 coefficient matrix
$\textbf{A}_k=\begin{bmatrix}
a_{xx,k} & a_{xy,k} \\
a_{yx,k} & a_{yy,k} \\
\end{bmatrix}$ quantifying the interactions within and between the two processes at lag $k$.

\subsection{Restricted model for Granger Causality and Granger Isolation}
Given the two stochastic processes $X$ and $Y$, the concept of Granger Causality (GC) is formalized quantifying the improvement in predictability that the past states of a putative driver process (say $X$) bring to the present state of the target process (say $Y$) above and beyond the predictability brought by the past states of the target itself \cite{granger1969investigating}. To implement this concept in the context of linear regression models, the present state of the target, $Y_n$, is described first from the past of both $X$ and $Y$ (respectively, $\textbf{X}_{n}^p=[{X}_{n-1} ... {X}_{n-p}]^\intercal$ and $\textbf{Y}_{n}^p=[{Y}_{n-1} ... {Y}_{n-p}]^\intercal$) through the so-called \textit{full} model (\ref{Y_ARX}), and then from the past of $Y$ only through the \textit{restricted} AR model
\begin{equation}
Y_n=\sum_{k=1}^\infty b_{yy,k}Y_{n-k}+U_{y|y,n},
\label{Y_Y}
\end{equation}
where $b_{yy,k}$ are autoregressive coefficients weighing the past samples of $Y$ and $U_{y|y,n}$ is a white noise process with variance $\sigma_{y|y}^2$. Note that the order of the restricted AR model is theoretically infinite \cite{barnett2015granger}; in practice, the model is implemented using $q$ lags, with $q$ sufficiently large.
Combining (\ref{Y_Y}) and (\ref{X_ARX}), a new bivariate AR model is designed where $Y_n$ is described as a function of $\textbf{Y}_{n}^q=[{Y}_{n-1} ... {Y}_{n-q}]^\intercal$ only, while the regression on $X_n$ remains unchanged:
\begin{equation}
\begin{aligned}
&X_n=\sum_{k=1}^p a_{xx,k}X_{n-k}+a_{xy,k}Y_{n-k}+U_{x|xy,n} \\
&Y_n=\sum_{k=1}^q b_{yy,k}Y_{n-k}+U_{y|y,n}.
\label{2ARmodel_GC}
\end{aligned}
\end{equation}
The coefficient matrix is here written as $\textbf{A}_{k}^Y=\begin{bmatrix}
a_{xx,k} & a_{xy,k} \\
0 & b_{yy,k} \\
\end{bmatrix}$, and quantifies the interactions within and between the two processes at lag $k$ when $Y$ is described through the AR restricted model obtained by truncating at lag $q$ the description of the past of $Y$ (\ref{Y_Y}).
Note that the measure of Granger Isolation (GI) is also derived from (\ref{2ARmodel_GC}), quantifying the degree of isolation of the putative target process $Y$ with respect to the causal effects brought by the putative driver process $X$.

\subsection{Restricted model for Granger Autonomy}
In analogy with GC, the concept of Granger Autonomy (GA) is formalized for a bivariate process assessing the predictability improvement brought to the present state of the target $Y$ by its own past states above and beyond the predictability brought by the past states of the driver $X$ \cite{seth2010measuring}. Operationally, GA is quantified comparing the \textit{full} model on $Y_n$ (\ref{Y_ARX}) with a \textit{restricted} cross-regressive (X) model whereby $Y_n$ is described only from the whole past of $X$, being the order of the X model theoretically infinite:
\begin{equation}
Y_n=\sum_{k=1}^\infty b_{yx,k}X_{n-k}+U_{y|x,n}
\label{Y_X},
\end{equation}
where $b_{yx,k}$ are cross-regression coefficients weighing the past samples of $X$ and $U_{y|x,n}$ is a noise process with variance $\sigma_{y|x}^2$. 
Thus, combining (\ref{X_ARX}) and (\ref{Y_X}) where the order is approximated with $q$ lags, a new bivariate AR model is designed where $Y_n$ is described as a function of $\textbf{X}_{n}^q=[{X}_{n-1} ... {X}_{n-q}]^\intercal$ only, while the regression on $X_n$ remains unchanged:
\begin{equation}
\begin{aligned}
&X_n=\sum_{k=1}^p a_{xx,k}X_{n-k}+a_{xy,k}Y_{n-k}+U_{x|xy,n} \\
&Y_n=\sum_{k=1}^q b_{yx,k}X_{n-k}+U_{y|x,n}.
\label{2ARmodel_GA}
\end{aligned}
\end{equation}
The coefficient matrix is here written as $\textbf{A}_{k}^X=\begin{bmatrix}
a_{xx,k} & a_{xy,k} \\
b_{yx,k} & 0 \\
\end{bmatrix}$, and quantifies the interactions within and between the two processes at lag $k$ when $Y$ is described through the X restricted model obtained by truncating at lag $q$ the description of the past of $X$ (\ref{Y_X}).

\subsection{Identification of restricted model parameters}
The restricted model coefficients, $b_{yy,k}$ and $b_{yx,k}$, and the variance of the AR and X model residuals, $\sigma_{y|y}^2$ and $\sigma_{y|x}^2$, appearing in (\ref{Y_Y}) and (\ref{Y_X}) respectively, can be identified starting from the covariance and cross-covariance matrices between the present and past variables of the two scalar processes $X$ and $Y$.
For jointly Gaussian processes, these matrices contain as scalar elements the covariance between two time-lagged variables taken from the processes $X$ and $Y$, which in turn appear as elements of the 2x2 autocovariance of the whole observed 2-dimensional process $\textbf{S}_n$, defined at each lag $k\geq{0}$ as $\mathbf{\Gamma}_k=\mathbb{E}[\textbf{S}_{n} \textbf{S}_{n-k}^\intercal]$, where $\mathbb{E}[\cdot]$ is the expectation operator. In this study, we follow the procedure described in \cite{faes2017information, faes2015information}, which exploits the possibility to compute $\mathbf{\Gamma}_k$ from the parameters of the bivariate AR formulation (\ref{2ARmodel}) of the process $\textbf{S}_n$ via the well-known Yule–Walker equations:
\begin{equation}
\mathbf{\Gamma}_k=\sum_{l=1}^p \textbf{A}_{l}\mathbf{\Gamma}_{k-l}+\delta_{k0}\mathbf{\Sigma}, \label{YWeq}
\end{equation}
where $\delta_{k0}$ is the Kronecher product. In order to solve this equation for $\mathbf{\Gamma}_k$, with $k=0,...,p-1$, we first express the bivariate AR model (\ref{2ARmodel}) in compact form as $\mathbf{\psi}_n=\textbf{A}\mathbf{\psi}_{n-1}+\textbf{E}_n$, where:
\begin{equation}\begin{split}
&\mathbf{\psi}_n=[\textbf{S}_{n}^\intercal \textbf{S}_{n-1}^\intercal ... \textbf{S}_{n-p+1}^\intercal]^\intercal; \\
&\textbf{A}=\begin{bmatrix}
\textbf{A}_1 & \dots & \textbf{A}_{p-1} & \textbf{A}_{p} \\
\textbf{I} & \dots & \textbf{0} & \textbf{0} \\
\vdots & \ddots & \vdots & \vdots \\
\textbf{0} & \dots & \textbf{I} & \textbf{0} \\
\end{bmatrix}; \\
&\textbf{E}_n=[\textbf{U}_{n}^\intercal \textbf{0}_{1\times{2(p-1)}}]^\intercal.
\label{YWparameters}
\end{split}\end{equation}
Then, the $2p \times 2p$ covariance matrix of $\mathbf{\psi}_n$, which is defined as $\mathbf{\Psi}=\mathbb{E}[\mathbf{\psi}_{n} \mathbf{\psi}_{n}^\intercal]$ and has the form
\begin{equation}
\mathbf{\Psi}=\begin{bmatrix}
\mathbf{\Gamma}_{0} & \mathbf{\Gamma}_{1} & \dots & \mathbf{\Gamma}_{p-1} \\
\mathbf{\Gamma}_{1}^\intercal & \mathbf{\Gamma}_{0} & \dots & \mathbf{\Gamma}_{p-2} \\
\vdots & \vdots & \ddots & \vdots \\
\mathbf{\Gamma}_{p-1}^\intercal & \mathbf{\Gamma}_{p-2}^\intercal & \dots & \mathbf{\Gamma}_{0} \\
\end{bmatrix}
\label{YWcovmatrix},
\end{equation}
can be expressed as $\mathbf{\Psi}=\textbf{A}\mathbf{\Psi}{\textbf{A}^\intercal}+\mathbf{\Xi}$ where $\mathbf{\Xi}=\mathbb{E}[\textbf{E}_{n} \textbf{E}_{n}^\intercal]$ is the $2p \times 2p$ covariance of $\textbf{E}_n$. This last equation is a discrete-time Lyapunov equation, which can be solved for $\mathbf{\Psi}$ yielding the autocovariance matrices $\mathbf{\Gamma}_{0}, ... ,\mathbf{\Gamma}_{p-1}$ \cite{faes2015information}. Finally, the autocovariance can be calculated recursively for any lag $k\geq{p}$ by repeatedly applying Yule-Walker equations (\ref{YWeq}) up to the desired lag $q$, starting from the parameters of the bivariate AR representation of the observed Gaussian vector process in (\ref{2ARmodel}).
In this work, we selected $q=20$.

To summarize, the procedure described above is based first on computing the autocovariance sequence of the bivariate process (\ref{2ARmodel}) from its AR parameters ($\textbf{A}_l$, with $l=1,...,p$, and $\mathbf{\Sigma}$), which are previously identified through the vector least square approach, and then on rearranging the elements of the autocovariance matrices for building the auto- and cross-covariances to be used in the computation of the AR parameters $b_{yy,k}$ and $\sigma_{y|y}^2$, and of the X parameters $b_{yx,k}$ and $\sigma_{y|x}^2$, appearing in (\ref{Y_Y}) and (\ref{Y_X}) respectively.
The identification of the restricted models in (\ref{2ARmodel_GC}) and (\ref{2ARmodel_GA}), as well as of the full model in (\ref{2ARmodel}),
will be then followed by their representation in the $Z$ domain and the computation of frequency-specific measures of GC, GI and GA.

\subsubsection{Restricted model for GC and GI}
The AR model (\ref{Y_Y}) can be written in compact form as $Y_n=\textbf{B}_{yy}\textbf{Y}_{n}^q+U_{y|y,n}$, where $\textbf{B}_{yy}=[b_{yy,1} ... b_{yy,q}]$ is the vector collecting all coefficients. From this representation, taking the expectation $\mathbb{E}[Y_n \textbf{Y}_{n}^{q^\intercal}]$ and solving for $\textbf{B}_{yy}$ yields:
\begin{equation}
    \textbf{B}_{yy}=\mathbf{\Sigma}_{Y_{n},Y_{n}^q} \cdot \mathbf{\Sigma}_{Y_{n}^q}^{-1} 
    \label{coeff_restricted_GC},
\end{equation}
where $\mathbf{\Sigma}_{Y_{n}^q}$ is the $q \times q$ autocovariance matrix of $\textbf{Y}_{n}^q$ defined as $\mathbf{\Sigma}_{Y_{n}^q}=\mathbb{E}[\textbf{Y}_{n}^q \textbf{Y}_{n}^{q^\intercal}]$, while $\mathbf{\Sigma}_{Y_{n},Y_{n}^q}$ is the $1 \times q$ cross-covariance matrix of $Y_n$ and $\textbf{Y}_{n}^q$, defined as $\mathbf{\Sigma}_{Y_{n}Y_{n}^q}=\mathbb{E}[Y_n\textbf{Y}_{n}^{q^\intercal}]$.
The matrices $\mathbf{\Sigma}_{Y_{n}^q}$ and $\mathbf{\Sigma}_{Y_nY_{n}^q}$ are extracted from $\mathbf{\Gamma}_{k}$.
Then, the variance of the AR residuals $\sigma_{y|y}^2$ in (\ref{Y_Y}) is computed as:
\begin{equation}
    \sigma_{y|y}^2=\sigma_{y}^2 - \mathbf{\Sigma}_{Y_n,Y_{n}^q} \cdot \mathbf{\Sigma}_{Y_{n}^p}^{-\intercal} \cdot \mathbf{\Sigma}_{Y_n,Y_{n}^q}^\intercal 
    \label{variance_restricted_GC},
\end{equation}
where $\sigma_{y}^2=\mathbb{E}[Y_n^2]$ is the variance of $Y$.

\subsubsection{Restricted model for GA}
Analogously to (\ref{coeff_restricted_GC}), the X model coefficients $b_{yx,k}$ in (\ref{Y_X}) are computed as:
\begin{equation}
    \textbf{B}_{yx}=\mathbf{\Sigma}_{Y_n,X_{n}^q} \cdot \mathbf{\Sigma}_{X_{n}^q}^{-1} 
    \label{coeff_restricted_GA},
\end{equation}
where $\textbf{B}_{yx}=[b_{yx,1} ... b_{yx,p}]$, $\mathbf{\Sigma}_{X_{n}^q}$ is the $q \times q$ autocovariance matrix of $\textbf{X}_{n}^q$ defined as $\mathbf{\Sigma}_{X_{n}^q}=\mathbb{E}[\textbf{X}_{n}^q \textbf{X}_{n}^{q^\intercal}]$, while $\mathbf{\Sigma}_{Y_n,X_{n}^q}$ is the cross-covariance matrix of $Y_n$ and $\textbf{X}_{n}^q$, defined as $\mathbf{\Sigma}_{Y_{n},X_{n}^q}=\mathbb{E}[Y_n \textbf{X}_{n}^{q^\intercal}]$.
The matrices $\mathbf{\Sigma}_{X_{n}^q}$ and $\mathbf{\Sigma}_{Y_n,X_{n}^q}$ are extracted from $\mathbf{\Gamma}_{k}$.
Then, the variance of the X residuals $\sigma_{y|x}^2$ in (\ref{Y_X}) is computed as:
\begin{equation}
    \sigma_{y|x}^2=\sigma_{y}^2 - \mathbf{\Sigma}_{Y_n,X_{n}^q} \cdot \mathbf{\Sigma}_{X_{n}^q}^{-\intercal} \cdot \mathbf{\Sigma}_{Y_n,X_{n}^q}^\intercal 
    \label{variance_restricted_GA}.
\end{equation}

\section{Interpretation of Granger Isolation and Granger Autonomy}
In this section, we elaborate on the definition  of the proposed measures of Granger Isolation and Granger Autonomy in the frequency domain, and provide further interpretation analyzing them in theoretical examples of coupled processes.

As pointed out in the main text, the concept of GA reflects the internal information in the target, being always zero in the absence of internal dynamics. Consequently, finding it higher than zero means that autodependency effects take place in the target process itself. The underlying idea is based on the consideration that a target system is autonomous if it is not controlled by external influences (i.e., driver processes) but self-determines its states \cite{seth2010measuring, boden1996philosophy, bertschinger2008autonomy}. In this sense, under the assumptions of Gaussianity and linearity \cite{barnett2009granger}, the concept of GA is formalized quantifying the improvement in predictability that the past states of the target bring to the present state of the target itself above and beyond the predictability brought by the past states of the driver \cite{seth2010measuring, faes2015information, porta2015conditional}. This definition relies on the identification of full and restricted AR models, whereby the partial variance of the present state of the target conditioned on the knowledge of the past states of the driver (\ref{Y_X}) is compared to the partial variance of the present state of the target conditioned on the knowledge of the past states of both the driver and the target (\ref{Y_ARX}) \cite{seth2010measuring, faes2015information, porta2015conditional}. If the latter is reduced with respect to the former, the GA measure increases in accordance with the predictability improvement brought by the past of the target system.
On the other hand, the GI is derived directly from the spectrum decomposition of the target process, expressed as the sum of causal and non-causal contributions, considering the non-causal contribution; as such, it quantifies that part of the target spectrum which cannot be explained by the driver, thereby informing about how much the target dynamics are “isolated” with respect to the driver process.

\begin{figure} [h!]
    \centering
    \includegraphics[scale=0.48]{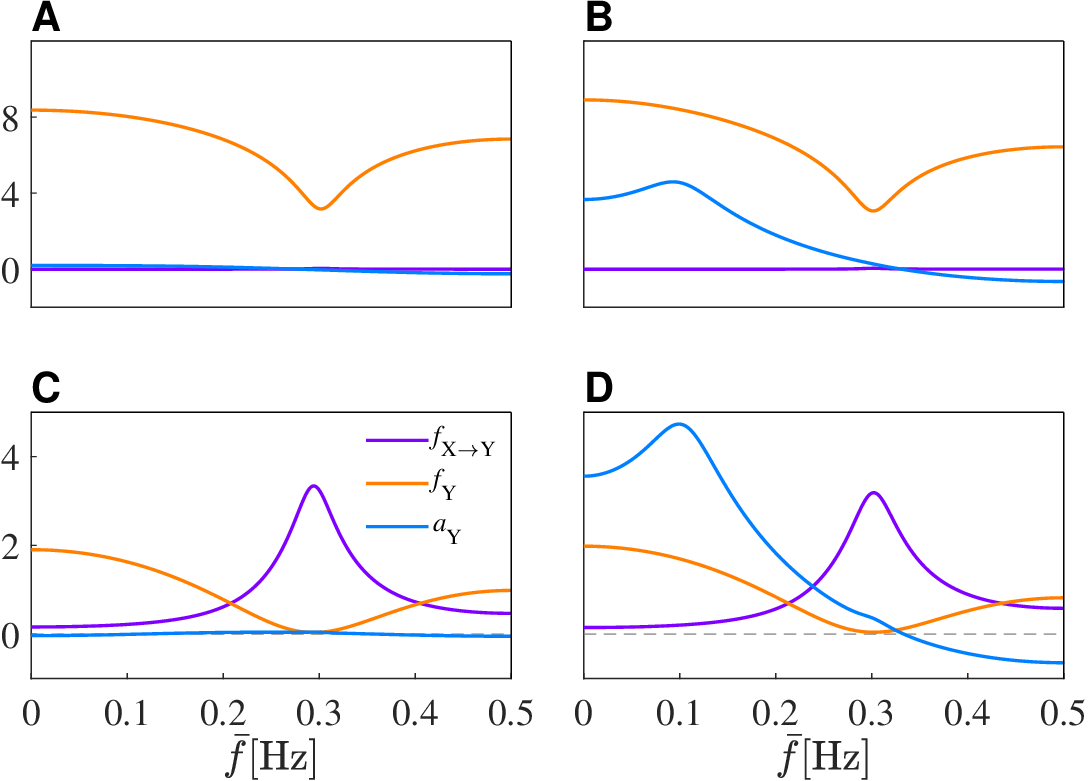}
\caption{Spectral behaviour of measures of GC ($f_{X\to{Y}}$, purple line), GI ($f_{Y}$, orange line) and GA ($a_{Y}$, light blue line) in different simulation settings. In each of these settings, the driver process $X$ is simulated as an AR process with an autonomous oscillation at $0.3$ Hz. (\textbf{A}) setting (i), isolated process without autonomous dynamics: the target is a Gaussian white noise with zero mean and unit variance. (\textbf{B}) setting (ii), isolated process with autonomous dynamics: the target is a process with self-dependencies featuring a stochastic oscillation at $0.1$ Hz. (\textbf{C}) setting (iii), non-isolated process without autonomous dynamics: the target is a process $Y$ with no self-dependencies but causally driven by $X$. (\textbf{D}) setting (iv), non-isolated process with autonomous dynamics: the target is a process with self-dependencies featuring an oscillation at $0.1$ Hz, and is causally driven by $X$.}
\label{simu_difference_autonomy_isolation}
\end{figure}

To visualize these concepts, here we study the behavior of the measures of GC, GI and GA using AR processes simulated in four different settings.
Specifically, we consider a bivariate AR process $\textbf{S}_n=[X_n Y_n]^\intercal$ whose dynamics are described by the following linear AR model:
\begin{equation}
\begin{aligned}
X_n&=a_{x,1}X_{n-1}+a_{x,2}X_{n-2}+U_n \\
Y_n&=a_{y,1}Y_{n-1}+a_{y,2}Y_{n-2}-c X_{n-1}+V_n
\label{sim_GI_GA}
\end{aligned}
\end{equation}
where $U$ and $V$ are independent Gaussian white noises with zero mean and unit variance.
The observed driver $X$ is simulated as an AR process with an autonomous oscillation obtained placing a pair of complex-conjugate poles, with modulus $\rho_x=0.9$ and phase $2\pi{f_x}$, $f_x=0.3$ Hz, in the complex plane representation of the process \cite{faes2015information}, so that its autonomous dynamics are determined by the fixed coefficients $a_{x,1}= -0.556, a_{x,2}= –0.81$.
The autonomous dynamics of the target $Y$, which are determined by the coefficients $a_{y,1}, a_{y,2}$ related to the complex-conjugate poles with modulus $\rho_y$ and phase $2\pi{f_y}$, depend on the parameter $b$ in such a way that $\rho_y=b \cdot 0.8$ (with $f_y=0.1$ Hz).
Moreover, a causal interaction is set from $X$ to $Y$ at lag $k=1$, with strength modulated by the parameter $c$.
We consider four parameter configurations, in which the driver dynamics are fixed and the target process $Y$ is simulated as: (i) an isolated Gaussian white noise with zero mean and unit variance ($b=0$, $c=0$); (ii) an isolated AR process with an autonomous oscillation at $f_y=0.1$ Hz ($b=1$, $c=0$); (iii) an AR process with no autonomous oscillations but influenced by $X$ at lag $k=1$ ($b=0$, $c=1$); (iv) an AR process with an autonomous oscillation at $f_y=0.1$ Hz and causally driven by $X$ at lag $k=1$ ($b=1$, $c=1$).
For each setting, a pair of time series $Y$ and $X$ of length $N=500$ is generated from the AR model parameters feeding the model with white noise realizations.
Then, model identification is performed for the bivariate process $\textbf{S}_n$, setting the model order at the value $p=2$. The spectral profiles of the measures of GC, GI and GA resulting from the generated time series are reported in Fig. \ref{simu_difference_autonomy_isolation}.

Results confirm that GI and GA measures reflect different behaviours of the target system.
Fig. \ref{simu_difference_autonomy_isolation}A reports the results relevant to the setting (i); the absence of internal dynamics for $Y$ (i.e., AR model coefficients relating the history of the target to its present state) is reflected by the flat spectrum of $a_Y(\bar{f})$, while the spectral profile of the GI, together with the GC, ensures the “isolation" of the target with respect to the driver.
Indeed, while the GC is uniformly zero at all frequencies, thus reflecting the absence of causal interactions from $X$ to $Y$, the spectral profile of the GI is characterized by a reverse peak at the frequency of the autonomous oscillation of the driver.
These findings hold again in case of an AR target process $Y$ whose dynamics are only self-determined (Fig. \ref{simu_difference_autonomy_isolation}B, setting (ii)).
On the other hand, in case of a well-established causal interaction from $X$ to $Y$ (Fig. \ref{simu_difference_autonomy_isolation}C, setting (iii), and Fig. \ref{simu_difference_autonomy_isolation}D, setting (iv)), the GC shows a clear peak at the frequency of the autonomous oscillation of the driver, and the GI still owns a reverse peak at the same frequency.
This finding is quite interesting, since it emerges that the profile of the GI may be related to the time- and frequency-domain behaviour of the driver, which indeed remains unaltered in all the settings (AR process with an autonomous oscillation at $0.3$ Hz). In another simulation, which has not been shown here for brevity, we found that the spectral profile of the GI is flat and uniform at all frequencies when $X$ is an isolated Gaussian noise. This suggests that the reverse peak of the GI, when present, may give indications on the frequency-specific location of a possible driving effect reducing the degree of isolation of the target process. 
These results seem to confirm that the measure of GI has a dual role. First, it quantifies that part of the target spectrum which cannot be explained by the driver, i.e., how much the target dynamics are “isolated” with respect to the driver process. Moreover, it gives also an information on the frequency-specific location of the “isolation" phenomenon, thus allowing to identify the spectral bands where this is less/more accentuated. 
On the other hand, the spectral profile of the GA is in line with what we expect to find, i.e. a positive peak in presence of an autonomous oscillation of the target process whereby it is located (Fig. \ref{simu_difference_autonomy_isolation}B,D), or conversely a flat spectrum for $a_Y(\bar{f})$ whether the target has no internal dynamics.

In conclusion, the differences between the two measures come out clearly. The two concepts of “isolation" and “internal dynamics" reflect different mechanisms occurring in the target process, as one does not rule out the other. They can coexist in the same process (Fig. \ref{simu_difference_autonomy_isolation}D), exist independently of each other (Fig. \ref{simu_difference_autonomy_isolation}B,C), or not exist at all (Fig. \ref{simu_difference_autonomy_isolation}A). Still, it is fundamental to investigate the pairwise interactions between the target and the driver through the three discussed measures, in order to have a complete overview of how two processes interact and/or oscillate independently of each other.

\bibliographystyle{IEEEtran}